\documentclass[aps,twocolumn, superscriptaddress, amssymb, prb, floatfix]{revtex4-2}

\usepackage{hyperref}
\hypersetup{
    colorlinks=true,
    linkcolor=blue,
    filecolor=magenta,      
    urlcolor=cyan,
    pdfpagemode=FullScreen,
    }

\usepackage{amsmath}
\usepackage{graphicx}
\usepackage{dcolumn}
\usepackage{bm}

\usepackage[utf8]{inputenc}
\usepackage[T1]{fontenc}
\usepackage{mathptmx}
\usepackage[dvipsnames]{xcolor}
\usepackage{geometry}
\usepackage[explicit]{titlesec}

\usepackage{xcolor}
\usepackage[final]{changes}
\definechangesauthor[color=red, name={Paulo Santos}]{PVS}
\definechangesauthor[color=teal, name={Alessandro Pitanti}]{AP}

\definechangesauthor[color=green, name={Madeleine Msall}]{MM}

\newcommand{\dPVS}[1]{\deleted[id=PVS]{#1}}
\newcommand{\cPVS}[1]{\comment[id=PVS]{ \small #1}}

\newcommand{\rAP}[2]{\replaced[id=AP]{#1}{#2}}
\newcommand{\aAP}[1]{\added[id=AP]{#1}}

\newcommand{\cAP}[1]{\comment[id=AP]{ \small #1}}

\newcommand{\cMM}[1]{\comment[id=MM]{ \small #1}}

\newcommand{\beginsupplement}{%
        \setcounter{table}{0}
        \renewcommand{\thetable}{SM\arabic{table}}%
        \setcounter{figure}{0}
        \renewcommand{\thefigure}{SM\arabic{figure}}%
        \setcounter{equation}{0}
        \renewcommand{\theequation}{SM\arabic{equation}}%
        \setcounter{section}{0}
        \renewcommand{\thesection}{SM\arabic{section}}%
     }

\newcommand{\fSAW}[0]{f_\mathrm{SAW}}         

\newcommand{\lSAW}[0]{\lambda_\mathrm{SAW}}
\newcommand{\vSAW}[0]{v_\mathrm{SAW}}
\newcommand{\keff}[0]{k^2_\mathrm{SAW}}

\newcommand{\fRF}[0]{f_\mathrm{RF}}           

\newcommand{\fBAW}[0]{f_\mathrm{BAW}}               

\newcommand{\rfilm}[0]{r_\mathrm{film}}
\newcommand{\dfilm}[0]{d_\mathrm{film}}
\newcommand{\Vrf}[0]{V_\mathrm{rf}}

\begin{document}



\title{GHz helical acoustic drum modes on a chip}

\author{N. Ashurbekov}
\affiliation{Paul-Drude-Institut f{\"u}r Festk{\"o}rperelektronik, Leibniz-Institut im Forschungsverbund Berlin e. V., Hausvogteiplatz 5-7, 10117 Berlin, Germany}
\author{I. dePedro-Embid}
\affiliation{Paul-Drude-Institut f{\"u}r Festk{\"o}rperelektronik, Leibniz-Institut im Forschungsverbund Berlin e. V., Hausvogteiplatz 5-7, 10117 Berlin, Germany}
\author{A. Pitanti}
\affiliation{Paul-Drude-Institut f{\"u}r Festk{\"o}rperelektronik, Leibniz-Institut im Forschungsverbund Berlin e. V., Hausvogteiplatz 5-7, 10117 Berlin, Germany}
\affiliation{University of Pisa, Dipartimento di Fisica E. Fermi,
largo Bruno Pontecorvo 3, Pisa 56127, Italy}
\altaffiliation[Also at ]{CNR - Istituto Nanoscienze, Laboratorio NEST 
Piazza San Silvestro 12, Pisa 56127, Italy}
\author{M. Msall}
\affiliation{Paul-Drude-Institut f{\"u}r Festk{\"o}rperelektronik, Leibniz-Institut im Forschungsverbund Berlin e. V., Hausvogteiplatz 5-7, 10117 Berlin, Germany}
\affiliation{Department of Physics and Astronomy, Bowdoin College, Brunswick, Maine 04011, USA}
%
%
\author{P. V. Santos}
\affiliation{Paul-Drude-Institut f{\"u}r Festk{\"o}rperelektronik, Leibniz-Institut im Forschungsverbund Berlin e. V., Hausvogteiplatz 5-7, 10117 Berlin, Germany}
\email[corresponding author: ]{santos@pdi-berlin.de}

\date{\today}


\begin{abstract}

On-chip laterally confined GHz acoustic modes with tunable helicity open the way for advanced optomechanical functionalities. Here, we demonstrate a novel concept for the implementation of these functionalites through the electrical excitation of GHz membrane-like drum modes. Our concept relies on the strong dependence of the frequency spectrum of Lamb acoustic modes on the thickness of the propagating medium. Lamb modes generated by a piezoelectric resonator in a thicker (or thinner) substrate region remain confined in this region via reflections at the lateral boundaries with thinner (thicker) substrate regions. If the generation region is disk-shaped, the lateral reflections form drum-like modes, which we experimentally confirm by radio-frequency spectroscopy as well as by maps of the surface displacements. Furthermore, we show that an array of sector-shaped piezoelectric transducers powered with appropriate radio-frequency phases creates helical modes with tunable helicity, which can then be transferred to an optical beam. 
Our analytical and finite-element  models  yield useful insights into the acoustic coupling of bulk and surface modes that can guide adaptation to other material systems. The acoustic drum modes thus provide a flexible platform for acousto-optical chiral functionalities in the GHz frequency range.  
%

%
\end{abstract}

\maketitle 



\section{Introduction}
\label{Introduction}

Thin acoustic membranes form a powerful platform for sensing and transduction by combining a very small mass with record-high quality factors  (Q)~\cite{Yuan_APL107_263501_15,Hofer_PRA82_31804_10}. 
The high Q's arise from a large acoustic impedance mismatch with the clamping medium and enable large mechanical deformations under an external excitation. 
Thin membranes, as well as suspended beam structures, can be conveniently defined using conventional lithography and under-etching procedures. These structures can be shaped into two-dimensional (2D) phononic and/or photonic crystals with enhanced opto-mechanical coupling. When combined with the large deformations, they become  particularly interesting for ultra-sensitive sensing \cite{piller2022,vicarelli2022} as well as for the  control of opto-\cite{Kapfinger_NC6__15,Zanotto_AOM8_1901507_19,Pitanti_AOM12_24}  and electronic quantum excitations~\cite{Teufel_N471_204_11,Usami_NP8_168_12,Seis_NC13__22}. 

Recent demonstrations of membranes with  topological phononic structures~\cite{Xue_NRM7_974_22,Zhang_CP1__18} that support helical acoustic modes are exciting high interest for particle manipulation in liquids~\cite{Riaud_PRA7_24007_17}, negative and asymmetric acoustic reflection~\cite{PVS351}, non-reciprocal acoustic propagation~\cite{Fleury_NC6__15}, the creation of synthetic gauge fields~\cite{Mathew_NN15_198_20}, as well as for the  acousto-optical generation of chiral light beams with tunable orbital angular momentum\cite{Riaud_PRA4_34004_15,Riaud_PRA7_24007_17,Zanotto_AOM8_1901507_19,PVS367} for optical communication and advanced metrology~\cite{Cheng_LS&A14__25}. However, the small thicknesses and masses of such suspended structures, which ensure their high Q-factors, also limit the mechanical vibration frequencies. With a few exceptions, most of the membrane studies have so far addressed acoustic modes with relatively low frequencies (i.e., well below a GHz).

\begin{figure*}[tbhp]
	\centering
		\includegraphics[width=0.95\textwidth]{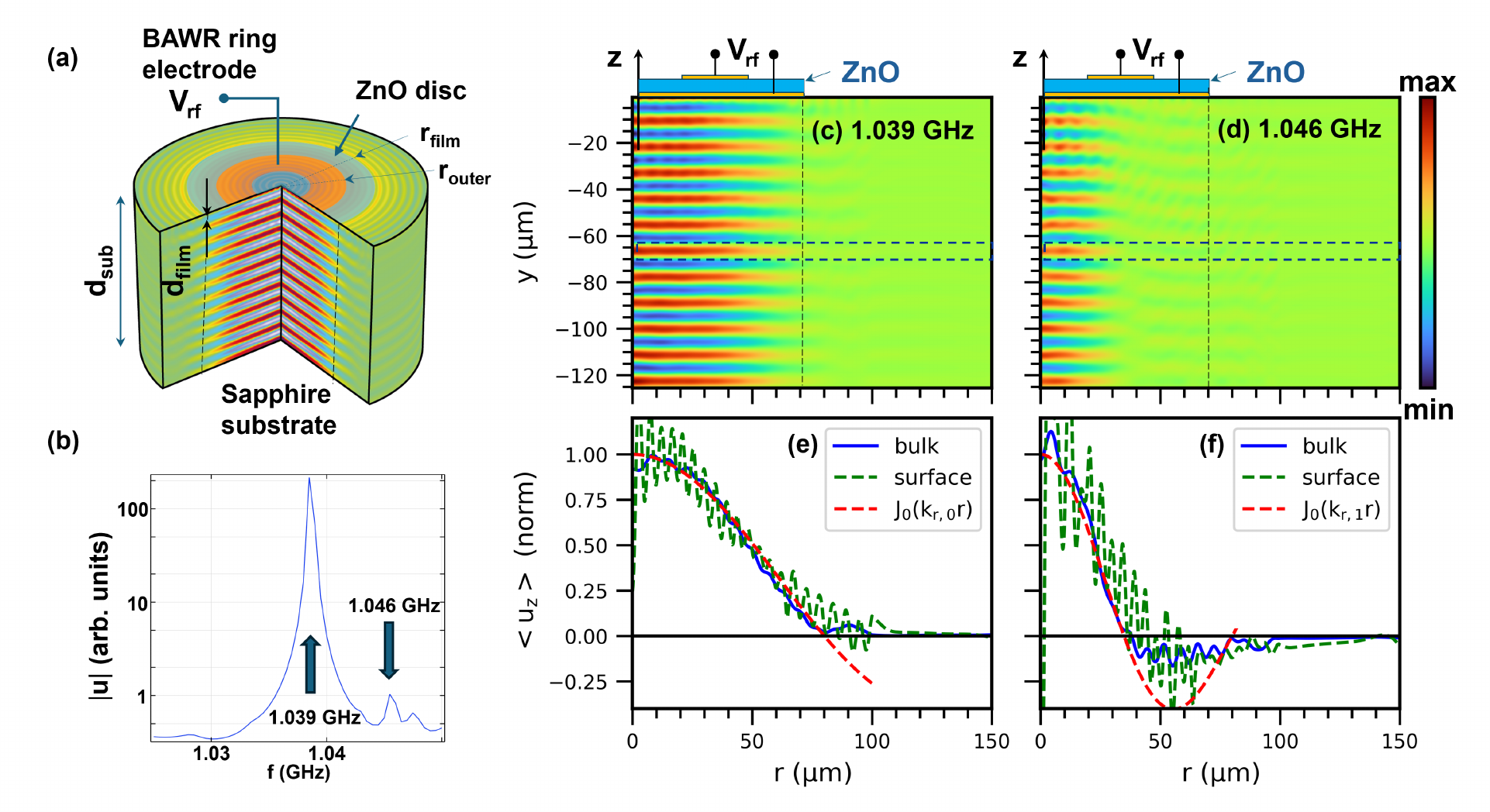}
\caption{ {\bf Laterally confined Lamb modes.}
  (a) 
  Finite-element (FEM) simulations of the acoustic displacement field ($\bf u$) excited by a ring-shaped piezoelectric  bulk acoustic wave resonator  (BAWR) on a sapphire substrate (see Methods for detail). The BAWR consists of a  piezoelectric ZnO layer of thickness $\dfilm=700$~nm shaped as a disk of radius of $\rfilm = 70~\mu$m and sandwiched between two thin metal contacts (cf.~upper part of panel c). The bottom contact is a $r_\mathrm{outer}=27~\mu$m while the top contact is a 30~nm-thick Al ring with inner and outer radii of $21.5~\mu$m and $r_\mathrm{outer}=27~\mu$m, respectively. The double-side polished c-plane sapphire substrate 
  (thickness $d_\mathrm{sub}=425~\mu$m)  
  forms Fabry-P\'{e}rot acoustic cavities with different lengths in the ZnO coated and uncoated regions, leading to different acoustic resonance frequencies in these two regions. The frequency mismatch laterally confine acoustic modes underneath the ZnO-coated region, which is demarked by vertical dashed lines in panels (a), (c) and (d). 
  (b) Calculated frequency ($\fRF$) spectrum of the average displacement amplitude $|\bf u|$ excited by the radio-frequency (rf) voltage ($\Vrf$) applied to the BAWR within a free spectral range of the acoustic cavity. 
  (c)-(d) Vertical displacement field ($u_z$) for the two resonances indicated by the thick arrows in panel (b), corresponding to modes confined underneath the ZnO layers. 
  (e)-(f) Radial dependence of the average acoustic amplitude $u_z$ integrated within the dashed rectangles (blue curves) and at the sample surface (green curves) in panels (c) and (d), respectively. The dashed red lines are Bessel-function fits to the profiles with different radial wavevectors $k_{r,m_r}$, ($m_r=1,2$) as discussed in Sec.~\ref{Radially confined Lamb modes}. The material parameters used in the finite element simulation are listed in  Table~\ref{appendix:pTable} of the Supplement (SM).
} 
\label{Fig1}
\end{figure*}

In this work, we first introduce  a novel concept for the piezoelectric excitation of  membrane-like helical drum modes  in the GHz range with tunable helicity on a bulk substrate  compatible with monolithic integration.  
The concept is illustrated by the finite-element (FEM) simulations displayed in Fig.~\ref{Fig1}(a) (see Methods for details). Here, acoustic waves are electrically generated by a ring-shaped bulk acoustic resonator (BAWR) consisting of two ring-shaped metal contacts (orange ring) sandwiching a disk-shaped piezoelectric film (ZnO, gray disk)  [cf.~\rAP{Fig.~\ref{Fig1}(a)}{upper part of panel (b)}].  
%
The finite radial dimension of the BAWR imparts a non-vanishing radial wave vector $k_r$ to the modes, which evolve into Lamb modes as they specularly reflect at the (polished) top and bottom sample surfaces. The volume between these surfaces forms  Fabry-P\'{e}rot cavities for the  Lamb modes with notably different resonance frequencies in the film-coated and uncoated areas. As a consequence, modes  resonantly excited by the BAWR underneath the circular film do not normally find matching states in the surrounding area. \aAP{Despite the homogeneous substrate material,} \rAP{and s}{S}imilarly to membrane vibrations, \rAP{the modes}{they} are reflected at the film boundaries and remain confined within this region [demarked by vertical dashed lines in Fig.~\ref{Fig1}(a)], thus forming  GHz drum-like modes with cylindrical symmetry.

We provide experimental evidence for the drum-like modes by directly mapping the acoustic  surface displacement with $\mu$m lateral resolution.
Furthermore, we also demonstrate  the excitation of helical drum modes with tunable orbital angular momentum (OAM) by an array of sector-shaped BAWRs powered with appropriate radio-frequency (rf) phases. 
The generation of helical modes using transducer arrays~\cite{Riaud_PRA4_34004_15} and spiral-like structures has recently been reported~\cite{PVS367}.
The approach introduced here extends these functionalities to the GHz range while providing in-situ control of the helicity polarity. 
The helical acoustic deformation of top sample surface is qualitatively similar to a reflective spiral phase plate \cite{Rumala_OL45_1555_20,PVS367}. Optical reflection on this surface can thus generate optical beams with orbital angular momentum modulated at GHz frequencies for the previously mentioned applications. \aAP{In a different regime, appropriate designs with low surface reflectivity favors the exploitation of acousto-optical coupling based on dynamic diffraction within the substrate material. Similarly to what done in optical fibers, one can engineer specific resonances between optical and acoustic wavelengths in order to create a dynamical Bragg pattern controlled by the strain wave via the optoelastic effect \cite{lu2021}.}
The experimental results are well-accounted for by finite-element and analytical models  that yield useful insights into the acoustic coupling of bulk and surface modes and  guide adaptation to other applications and material systems.

In the following sections, we first demonstrate the feasibility of generating GHz drum modes by presenting numerical simulations for the structure in Fig.~\ref{Fig1}(a) (Sec.~\ref{Radially confined Lamb modes}).  
The experimental work includes investigation of the electrical response of the structures (Sec.~\ref{Electrical response}), the optical mapping of the spatial distribution of the drum acoustic modes (Sec. ~\ref{Mapping of GHz drum modes}), and the excitation of helical drum modes (Sec.~\ref{Helical Drum Modes}).  Section~\ref{Discussions}  introduces an analytical model for the spatial distribution of the acoustic field, which provides insight into the nature of the observed modes, thus supporting our main conclusions summarized in Sec.~\ref{Conclusions}. 

\section{Results}
\label{Results}

\subsection{Radially confined Lamb modes}
\label{Radially confined Lamb modes}

\dPVS{
Due to the finite radial dimension, the modes generated by BAWRs have a non-vanishing radial wave vector and evolve into Lamb modes as they specularly reflect at the (polished) sample surfaces. The sample bulk forms a Fabry-P\'{e}rot cavity for the Lamb modes with notably different resonance frequencies in the ZnO-coated and uncoated areas. As a consequence, modes resonantly excited by the BAWR are radially  confined underneath the circular ZnO film.  The radial boundary condition is responsible for the GHz drum-like modes observed when mapping surface displacement. }
\dPVS{In ~Fig.~\ref{Fig1}(a) we show such an FEM simulation of such a confined Lamb wave on a c-oriented sapphire substrate.  
In this simulation the ZnO film and the sapphire substrate are isotropic, ignoring the in-plane acoustic anisotropy of the trigonal sapphire substrate (cf. Supplementary Sec.~\ref{appendix:Acoustic waves in ZnO-coated sapphire}). Under this assumption, the acoustic displacement field ${\bf u}$ can be conveniently determined in cylindrical coordinates using a two-dimensional model. 
}

Further analysis of the FEM simulation results leading to Fig.~\ref{Fig1}(a) yields interesting insights into the lateral confinement mechanisms leading to the GHz drum modes. 
We will present results for BAWRs based on a textured ZnO piezoelectric film on a c-plane sapphire substrate; the concept is, however, extendable to other material combinations.
Figure~\ref{Fig1}(b) shows the dependence of the field amplitude $|{\bf u}|$ averaged over the substrate depth  as a function of the frequency $f$ driving the BAWR. The full spectrum consists of a sequence of lines with a periodicity in frequency $\Delta f_L =v_{L}/(2 d)$, $d=(d_\mathrm{sub}+\dfilm)$, equal to the splitting between the  Fabry-P\'{e}rot acoustic modes confined  in-between the top and (perfectly reflecting) bottom surface, where $v_L$ is the longitudinal acoustic velocity in the substrate and the dimensions $d_\mathrm{sub}$ and $\dfilm$ are defined in the figure. The spectral section in Fig.~\ref{Fig1}(b) details the frequency region close to  one of the Fabry-P\'{e}rot main modes. Instead of a single resonance, the spectrum shows several closely spaced peaks, attributed to distinct acoustic modes confined in the cavity regions underneath the ZnO layer. 

Panels (c) and (d) of Fig.~\ref{Fig1} show maps of the vertical surface displacement field ($u_z$) for the  resonances  marked by arrows in panel (b). These maps show comparable profiles along the $\hat z$ directions but rather distinct radial patterns.  
Panels~\ref{Fig1}(e) and \ref{Fig1}(f) display radial $u_z$ profiles for these modes  averaged within the blue dashed rectangles displayed in Figs.~\ref{Fig1}(c) and  \ref{Fig1}(d), respectively. The green dashed lines display the corresponding profiles recorded near the surface. Although the piezoelectric excitation is limited to the narrow, ring-shaped area underneath the BAWR, the field profiles for both modes extend over the whole region underneath the ZnO layer and decay beyond it. Note, in particular, that the profiles smoothly extend to the non-excited region at the center of the structure. The decay beyond the ZnO \rAP{border}{edge} is attributed to the previously mentioned thickness discontinuity of the Fabri-P\'erot acoustic cavity at this border, which effectively back-reflects the Lamb modes, thus leading to the formation of drum-like vibrations underneath the ZnO film. In fact, as will be justified  in Sec.~\ref{Lamb-waves in cylindrical coordinates}, the radial mode profiles follow the expected  Bessel-like radial shape $J_0( k_{r,i})$ with different radial wave vectors $k_{r,i}$, $i=0,1, 2 ,\dots$ [dashed lines in Figs.~\ref{Fig1}(e) and \ref{Fig1}(f)].\cAP{Maybe it could be useful to add a sentence about the fact that surface profiles show many oscillations.}  

\begin{figure*}[tbhp]
	\centering
		\includegraphics[width=1\textwidth]{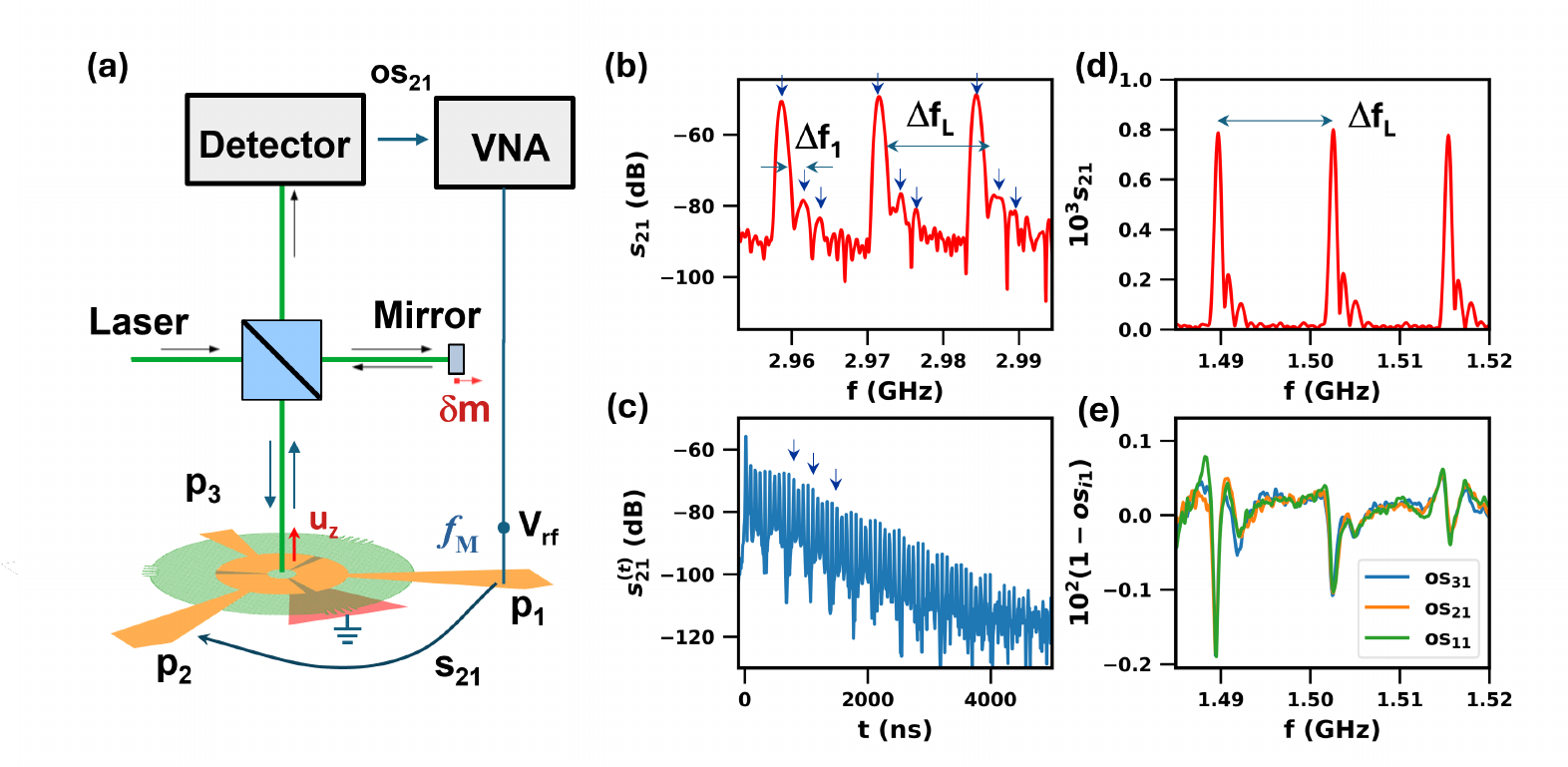}
\caption{
  {\bf Electrical excitation of Lamb acoustic modes.}
  (a) The ring-shaped BAWR has three sectors, which can be independently excited by radio-frequency fields. The electrical response is measured with a vector network analyzer (VNA). Maps of the surface displacement $u_z$ were recorded with phase resolution using a scanning Michelson interferometer  with spatial resolution of approx. $1~\mu$m (see Ref.~\onlinecite{PVS367} for  details).
  (b) Radio-frequency transmission ($s_{21}$ scattering parameter) between BAWR sectors $p_1$ and $p_2$ recorded in a frequency range around 3~GHz. $\Delta f_L=13.09$~MHz is the frequency difference between the modes. The arrows mark different resonances within a $\Delta  f_L$ period. 
  (c) Corresponding time-domain response recorded with a VNA with time-domain capabilities. The arrows show beats due to the interference of the different modes. 
  (d) $s_{21}$ spectrum in the 1.48-1.52~GHz frequency range and 
  (e) corresponding interferometric data recorded by exciting port $p_1$ and detecting the surface displacement $u_z$ at the center of BAWR $p_1$ (os$_{.11}$), $p_2$ (os$_{21}$), and $p_3$ (os$_{31}$). The os$_{i1}$ data were corrected to remove the unstructured background. 
} 
\label{Fig2}
\end{figure*}

\subsection{Electrical and optical responses}
\label{Electrical response}

The experiments were performed on samples with \rAP{a ring shaped BAWR divided in three independent sectors} {three sector-like BAWR} illustrated in Fig.~\ref{Fig2}(a). Figure~\ref{Fig2}(b) displays the rf-spectrum of the transmission parameter $s_{21}$ (corresponding to the acoustic transmission from sector $p_1$ to sector $p_2$) over a frequency band around 3 GHz.  The response consists of a series of pronounced spectral peaks separated by  
$\Delta f_L=13.09$~MHz corresponding to the spacing between the longitudinal resonances of the acoustic cavity defined by the substrate and overlaying ZnO layer obtained using $v_L$ value listed in Table~\ref{appendix:pTable}. The Q-factor of these electrical resonances, determined from the half-width of the $s_{21}$ peaks, is $3800 \pm 200 $. As in the simulations in Fig.~\ref{Fig1}(b), there is not just a single resonance per  $\Delta f_L$-period but rather a sequence of spectral lines (indicated by arrows in the figure). The frequency shift between the first two modes is approximately $\Delta f_1=2.53$~MHz$\ll \Delta f_L$. As will be justified in Sec.~\ref{Radially confined Lamb modes}, these resonances are attributed to modes radially confined  underneath the ZnO region.

Figure~\ref{Fig2}(c) shows a time-domain representation of $s_{21}$ obtained by a Fourier transformation of the frequency domain spectrum over a frequency range of 100~MHz. This temporal data, which approximates the transmission of a short pulse excitation, consists of a series of echoes displaced by $1/\Delta f_L=76.4$~ns and an envelope modulation with a periodicity of $\sim 1/\Delta f_1 = 400$~ns (the individual echoes are not resolved in the (long) time scale of the plot). 
The modulated envelope thus arises from the interference of  radially confined Lamb modes.\cMM{Need a bit more about why these timescales are thus ... perhaps estimate $2 pi/k_r$  and compared to radial dimension? wavelength - DONE with $1/\Delta f_1$}

The spectral shape with the  confined mode signatures are observed over a wide range of frequencies extending below a GHz, as illustrated by the $s_{21}$ spectrum in Fig.~\ref{Fig2}(d). 
The relationship between the electrical and optical responses is illustrated by the green curve (os$_{11}$) in Fig.~\ref{Fig2}(e). This spectrum was acquired by powering only $p_1$ and detecting \aAP{mechanical vibrations - see Methods -} at the center of BAWR $p_1$. The surface displacement ($u_z=os_{21}$) shows dips at the principal peaks of the electrical $s_{21}$ response of Fig.~\ref{Fig2}(d). 
As will be discussed in Sec.~\ref{Discussions},
small  differences (which also depend on frequency) between the optical and electrical responses arise from the different sensitivities of the two measurement techniques to Lamb waves with different radial wave vectors.

A further remarkable feature revealed by Fig.~\ref{Fig2}(e) is that by powering {\em only} port $p_1$ one measures essentially the same interferometric response at the center of the BAWRs at ports $p_1$ (os$_{11}$, green) $p_2$ (os$_{21}$, blue) and $p_3$  (os$_{31}$, red). 
As will be further elaborated in the next section, this behavior is attributed to the excitation by $p_1$ of drum-like modes of cylindrical symmetry in the region underneath the ZnO film.

\begin{figure*}[tbhp]
	\centering
		\includegraphics[width=0.9\textwidth, keepaspectratio=true]{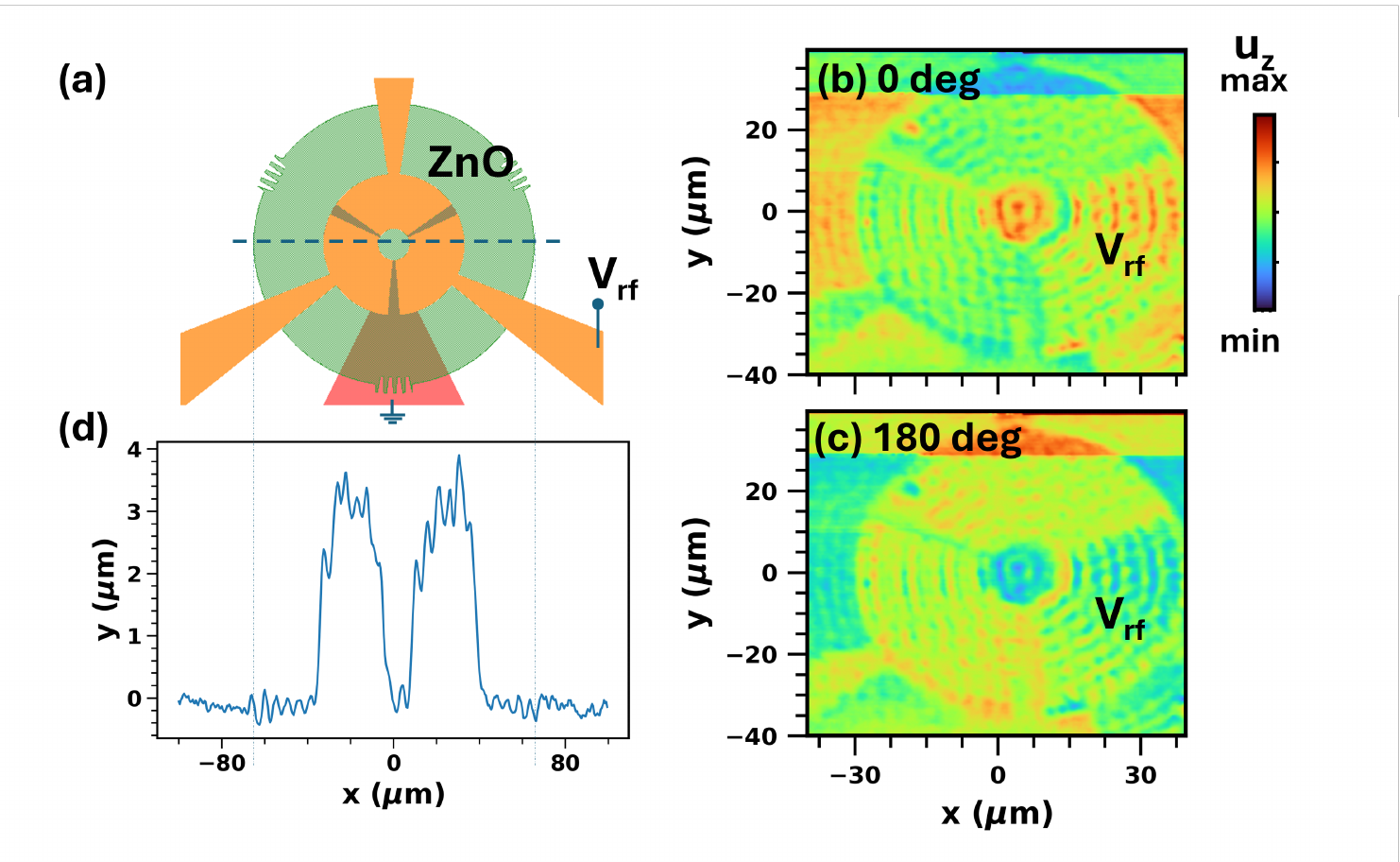}
\caption{
{\bf Interferometric mapping of drum modes.}
(a) Sample configuration with ring sector-shaped BAWRs with three three independent sections. 
(b)-(c) Acoustic maps of the surface displacement $u_z$ obtained by exciting {\em only} the lower right BAWR (marked as  $V_\mathrm{rf}$) at the frequency of 1.5033~GHz and recording the $u_z$-component in phase and 180$^\circ$ out-of-phase with the rf excitation. 
(d) Line profile of $u_z$ along the dashed line in panel (a) showing that the short-period acoustic vibrations (i.e., the ones with ring-shaped wave fronts in b and c) are confined within the ZnO island.
} 
\label{Fig3}
\end{figure*}


\subsection{Mapping of GHz drum modes}
\label{Mapping of GHz drum modes}

Interferometric maps that extend beyond the ZnO radial boundary [cf.~Fig.~\ref{Fig3}(a)] provide the key experimental evidence for the excitation of laterally confined drum-like modes. Figures~\ref{Fig3}(b) and \ref{Fig3}(c) compare surface maps for the surface displacement ($u_z$) components  in-phase and out-of-phase (i.e., with a phase shift $\phi_\mathrm{rf}=180^\circ$) with respect to the rf-drive, which was applied {\em only} to port $p_1$. 
Due to the lower optical reflectivity, the signal outside the BAWR contacts is much weaker than on the metal contact of the BAWRs, but also changes sign with the detection phase.

\begin{figure*}[tbhp]
	\centering
		\includegraphics[width=0.90\textwidth, keepaspectratio=true]{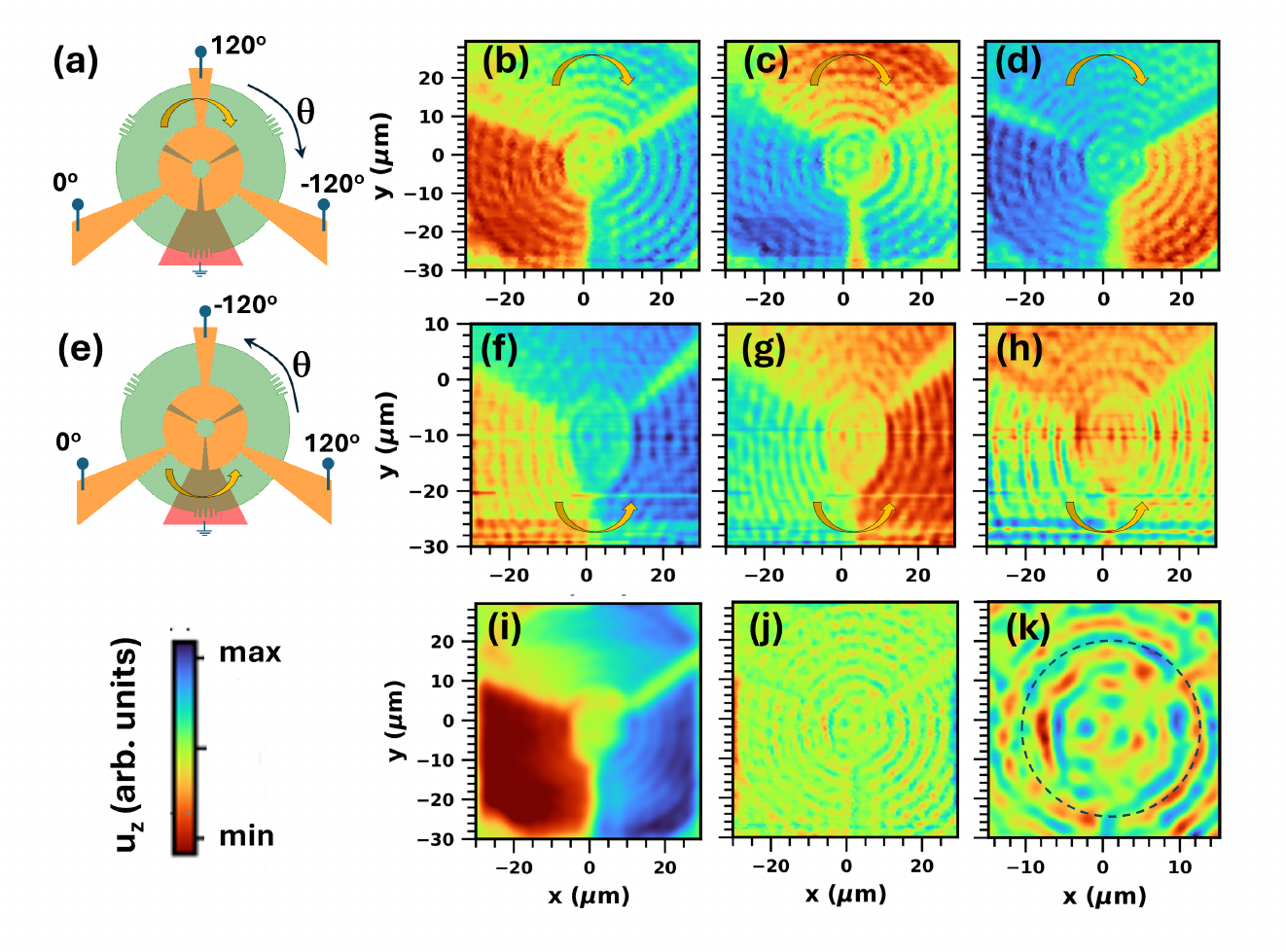}
\caption{{\bf Helical drum modes.}
 (a) Setup for the excitation of  helical waves by driving  the BAWR sectors with clock-wise increasing phases differing by 120$^\circ$.  (b)-(d) Profiles for the surface displacement field $u_z$ of helical modes at time instants corresponding to phases of (b) 0 deg, (c) 120 deg, and (d) 240 deg for $\fRF=1.5033$~GHz. (e-h) Corresponding results for the excitation of anti-clockwise helical waves. (i)-(j) Spatially filtered versions of panel (f) to review surface oscillations of long and short spatial periods, respectively. (k) Zoom view of the center region of panel (j) revealing the phase wrapping of the modes with short spatial periods. The  color code is the same for panels (b)-(h) and (i)-(j). In (k), it was expanded by a factor of two relative to (i)-(j).
 } 
\label{Fig4}
\end{figure*}

Interestingly, the recorded interferometric pattern is almost perfectly cylindrical with essentially the same amplitude on all sectors despite the excitation of just one of the sector-shaped BAWRs. 
The peculiar cylindrical shape, which is corroborated by the almost identical spectral shapes in Fig.~\ref{Fig2}(e), is attributed  to the preferential excitation of cylindrical drum-like modes arising from the radial confinement underneath the ZnO disk. 
Similar to a circularly clamped drum membrane, the amplitude of the Lamb modes excited by the BAWR is enhanced when their frequency matches a drum-like eigenstate: these modes are reflected at the borders of the ZnO disk to yield a cylindrical pattern with maximum displacement amplitude at the disk center. The shape of the sector-like BAWR thus plays only a minor role in the mode lateral profile.  The surface displacement should, therefore, have a Bessel-like profile $J_0(k_{r,m_r} r)$ with an effective radial wave vector $k_{r,m_r}$ given by: 

\begin{equation}
  k_{r,m_r} \approx  m_r\frac{\pi}{\rfilm}, \quad m_r=1,2,\dots.
  \label{Eqkr}
\end{equation}

\noindent where $\rfilm$ is the radius of the piezoelectric film. 

Modes with two distinct radial periodicities,  $\lambda_{r,m_r} = 2\pi/ k_{r,m_r} \approx 2 \rfilm/m_r$, can be identified in the experimental maps. The first yields strong fringes with a short radial periodicity and $m_r \sim 30$, similar to the short radial wavelength structure seen in the simulations of Fig.~\ref{Fig1}(e) and \ref{Fig1}(f) as well as in the spatial maps of Fig.~\ref{Fig3}. For the data in Fig.~\ref{Fig3} recorded for $f=1.50$~GHz, this component has $\lambda_{r,m_r} = 4.3 \pm 0.2 \mu m_r$. This together with several other experimentally observed modes of the same type are indicted by red asterisks in the prediced Lamb-wave dispersion displayed in Fig.~\ref{dispSymPlot}. As will be justified in Sec.~\ref{Lamb-waves in cylindrical coordinates}, they are attributed to the excitation of Lamb modes confined at the surface. 

The radial confinement of the short-period mode is clearly demonstrated in Fig.~\ref{Fig3}(d), which displays an interferometric line scan extending over the whole diameter of the ZnO disk [cf.~dashed line in Fig.~\ref{Fig3}(a)]. The larger $u_z$-oscillations on the BAWR contact areas arises again from  the higher reflectivity of these metal-coated areas compared to that of the ZnO-coated substrate. \cAP{Please check the y-axis label in Fig. 3 (d), I'm not sure it's correct.} 
The $u_z$-oscillations propagate over several tens of $\mu$m until they reach (and get reflected from) the border of the ZnO disk and decay away from it.

The second mode type has a much longer surface wavelength  $\lambda_{r,1} \sim 2 \rfilm$: it appears in Figs.~\ref{Fig3}(b) and \ref{Fig3}(c) as  a faint background superimposed on the fringes of the short-period mode [the latter is most clearly seen by spatially filtered the signal, cf.  Fig.~\ref{Fig4}(i)].
 The interference of the short- and long period modes builds the strong displacement antinode at the center of the pattern.
%
Due to the smaller surface displacement  amplitudes, the interferometric line profile of Fig.~\ref{Fig3}(d)  cannot clearly detect the radial confinement of the long period mode.

\subsection{ Helical Drum Modes}
\label{Helical Drum Modes}

The sector-like BAWR configuration can generate GHz helical waves when \aAP{all sectors are} powered with rf-voltages with phases $\phi_\mathrm{rf}$ increasing by 120~deg in the clock-wise direction [cf.~Fig.~\ref{Fig4}(a)].
Panels~\ref{Fig4}(b)-(d) show interferometric maps of the surface displacement recorded for this configuration at time instants corresponding to these phases.  
\cMM{DONE: I'm not sure how to best describe this phase advance.  I am assuming that the pattern was only measured once but that the figure shows the pattern with all the phases advanced by 0, 120 or 240 degrees.  We should take some care for how this is described in the figure caption.} 
Panels \ref{Fig4}(f)-(h) display the corresponding plots when the BAWRs are driven in a counter-clockwise direction [cf.~Fig.~\ref{Fig4}(e)].  The wave rotation can be readily identified by following the changes in the time-evolved interferometric maps around a clockwise or counter-clockwise path [e.g., as indicated by the curved arrows in Figs.~\ref{Fig4}(b) and (f)]. The observed phase wrapping results in the creation of vortices at the center of the pattern with topological charges $\ell=1$ and $\ell=-1$ for anti-clockwise and clockwise rotations, respectively.  Videos showing the full time evolution of the  clockwise and counter-clockwise rotating maps are included in Section~\ref{Displacement_Videos}.

The helical displacement profiles of Figs.~\ref{Fig4}(b)-(d) and \ref{Fig4}(f)-(h) are composed of vibrations with both long and short radial wave vectors.  In order to distinguish these two contributions, Figs.~\ref{Fig4}(i) and \ref{Fig4}(j) display the same data as in panel \ref{Fig4}(b) after spatial filtering wave components with long and short radial periods, respectively.  The long period mode shown in Fig.~\ref{Fig4}(i) has the strongest amplitude variation with the azymuthal angle $\theta$, forming a vortex at the center of the pattern. 

The interferometric maps [such as, e.g., the one in Fig.~\ref{Fig4}(i)] yield  the phase shift $\delta \phi(r,\theta) = 4 \pi |u_r(r,\theta, z=d/2)|/\lambda_{opt}$ of a focused optical beam scanned over the  sample surface $z=d/2$. The use of a focused optical beam enables reaching the required spatial ($\mu$m) and temporal (sub-ns) resolutions for mapping GHz acoustic fields. The mapping procedure is, otherwise, analogous to the holographic determination of the phase cross-section  of an extended optical beam  with sub-ns time resolution.
If the acoustic vortex region is illuminated by an extended optical beam (i.e., with dimensions compared to the one of the acoustic vortex), the phase cross-section of the reflected beam will then acquire the same phase pattern $\delta \phi(r,\theta)$, thus implying in the  transfer of angular momentum from the acoustic to the optical mode. \cAP{Well described ! :-)}

The short-period oscillations [cf.~Fig.~\ref{Fig4}(j)] have, in contrast, a weaker displacement amplitude with a smaller vortex at the center. Their helical character becomes evident when plotted over the smaller region shown in Fig.~\ref{Fig4}(k). By tracing a circular path (as shown by the dashed circle) around the vortex, one can easily confirm that the phase of the wavefronts shift by 120$^\circ$ between neighboring BAWRs with helicity polarity dictated by the rotation direction for phase increase. The complex shape of the vortex at the center arises from the discrete phase evolution with $\theta$.  A better approximation for a spiral wave front shape can be obtained by increasing the number of BAWR sectors.

\section{Discussions}
\label{Discussions}

We have demonstrated the piezoelectric  generation of drum-like Lamb waves propagating over several tens of $\mu$m, which can be radially confined using overlayers on the substrate. Furthermore, we have shown that  Lamb modes with positive and negative orbital angular momentum can be excited by an array of BAWRs driven by synchronized rf phases. The modes with acoustic angular momentum can be mapped with optical techniques, thus demonstrating the transfer of angular momentum to a light beam. 

\begin{figure*}[tbhp]
	\centering\includegraphics[width = 1\textwidth, keepaspectratio=true]{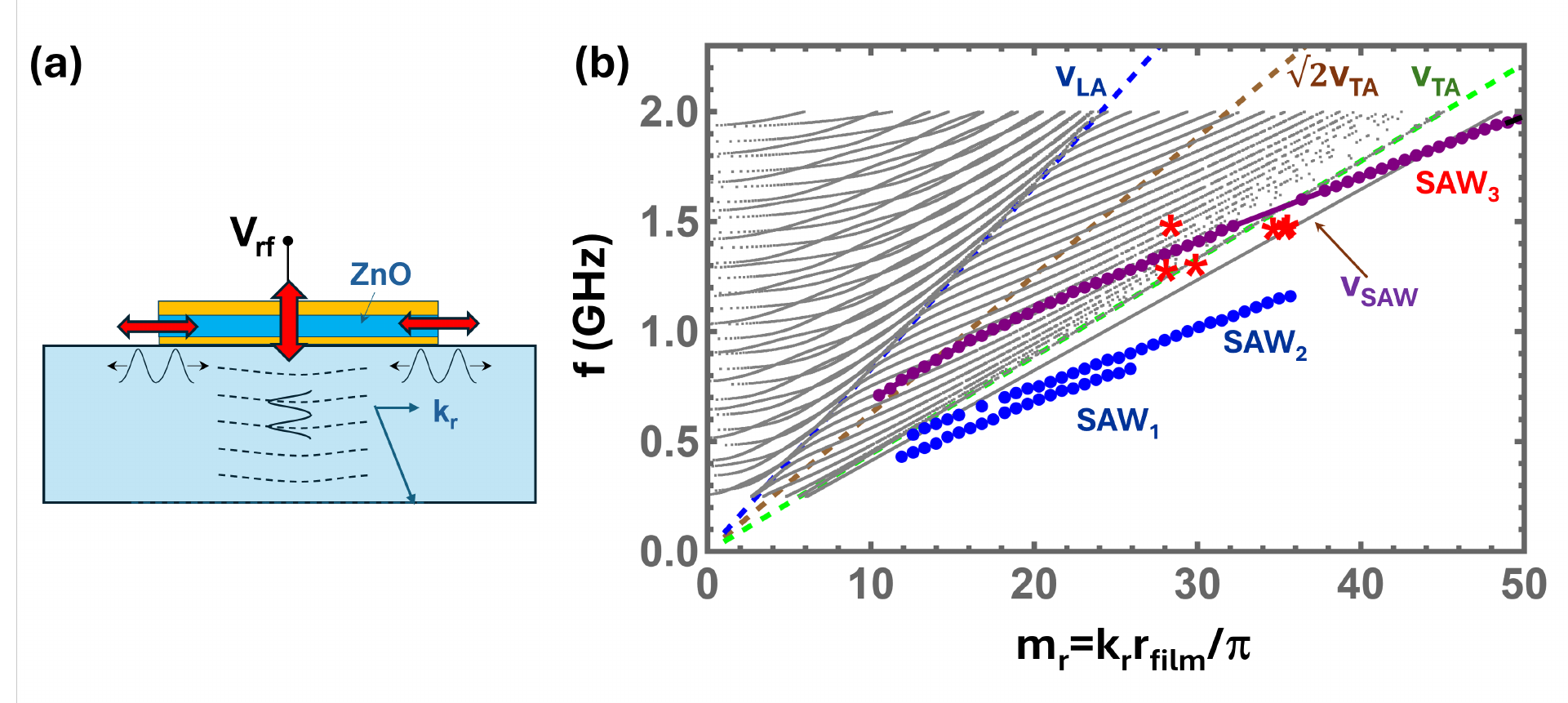}
 \caption{
  {\bf Dispersion of cylindrical  Lamb waves.} 
 Dispersion of $rz$-polarized Lamb waves in a cylindrical disk of sapphire of radius $\rfilm$ with winding number $n=0$ calculated assuming an isotropic elastic model for the sapphire substrate (cf.~Eq.~\ref{detSym}). The calculations were performed for a 425~$\mu$m-thick substrate: due to the high mode density, not all modes are shown.
The in-plane wave vector $k_r$ is normalized to $\pi/\rfilm$, where $\rfilm$ is the radius of the ZnO island.
The dashed lines indicate the dispersion of the pure longitudinal ($v_\mathrm{L}$, Eq.~\ref{eqkLkT}), transverse ($v_\mathrm{T}$, Eq.~\ref{eqkLkT}), Lam{\'e}  ($v_\mathrm{T}\sqrt{2}$, Eq.~\ref{EqvTsqrt}), and SAW modes ($\vSAW$, Eq.~\ref{EqSAW}) determined for the sapphire substrate.
 The blue and purple dots show the dispersion of SAW modes determined for a sapphire substrate coated with a 700~nm-thick ZnO film (cf.~Sec.~\ref{SAW modes in ZnO-coated c-plane sapphire}).
  The red asterisks are the corresponding experimental values for short period waves obtained from field maps (such as, e.g., those in Figs.~\ref{Fig3} and \ref{Fig4}). 
 }
\label{dispSymPlot}
\end{figure*}

Despite the monochromatic excitation, the BAWRs employed in the present studies generate Lamb waves with different radial wavevectors. Phenomenologically, the excitation of different radial wavevectors can be understood with the help of Fig.~\ref{dispSymPlot}(a). The rf-voltage applied to the BAWR electrodes induces an oscillating stress field in the piezoelectric film, which preferentially launches longitudinal modes towards the substrate. If the lateral size of the BAWR is small, it will, together with wave diffraction, impart a small radial  wave vector component  $k_{r,m_r}, m_r\sim 1$ to the launched wave  with amplitude dictated by the BAWR size (cf.~Eq.~\ref{Eqkr}). These waves then bounce back and forth at the top and bottom sample boundaries to form propagating Lamb modes with a small $k_{r,m_r}$. 

The deformation of the piezoelectric film along $z$ is, in addition, accompanied by a radial deformation as well as the creation of shear stresses at the lateral boundaries of the BAWR. These deformations can launch Lamb modes with a radial wave vector satisfying Eq.~\ref{Eqkr} for large $m_r$, corresponding to wavelengths considerably smaller that the BAWR dimensions. As will be discussed below, Lamb waves with large $k_r$ can become fully confined to the surface areas, eventually resulting in the the formation of surface acoustic waves (SAWs). 
Finally, if the BAWRs are driven at different phases as in Fig.~\ref{Fig4}, these phases will be imparted to the waves, leading to the formation of helical modes. 

The electrical (i.e., s-parameters) and optical (interferometry) field mapping techniques are expected to be most sensitive to SAW-like modes with  strong field localization near the surfaces. While interferometry can resolve the short wavelength modes, the electrical detection integrates the contributions of waves with radial wavelengths less than approximately twice the BAWR dimensions, thus making it primarily sensitive to Lamb modes with large radial wavelengths. These differences in detection sensitivity can qualitatively account for the different profile shapes in Fig.~\ref{Fig2}(d) and \ref{Fig2}(e). 
In the case of waves with non-zero angular momentum, the spatial filtering technique employed in Figs.~\ref{Fig4}(i) and \ref{Fig4}(j) shows that interferometry  can, in this case, detect the helical character of waves with both small and long spatial radial periods.  

In order to quantitatively analyze the experimental results, we first calculate in the next section the dispersion of cylindrical Lamb waves propagating in a disk. We then address the impact of the ZnO overlayer on these waves (Sec.~\ref{ZnO-induced surface modes}) as well as on their radial confinement (Sec.~\ref{Radially confined Lamb modes}). 

\vspace{1 cm}
\subsection{Dispersion of cylindrical Lamb-waves}
\label{Lamb-waves in cylindrical coordinates}

Lamb-waves in plates have been the subject of several studies (see, e.g., Refs.~\onlinecite{Lamb_Book_1910} and \onlinecite{Auld90a}) most of them addressing 1D Lamb modes propagating along a single direction. Only a few studies have considered the propagation of Lamb waves in the cylindrical geometry relevant to the present studies of Fig.~\ref{Fig1}(a)~\cite{Honarvar_IJSS44_5236_07}. The supplementary Sec~\ref{appendix:SI_Acoustic Modes} presents a detailed analytical calculation of these propagating in an elastically isotropic disk limited by free surfaces at $z=\pm d/2$. Here, we summarize the main results.


The relevant Lamb modes for our experimental configuration are those that can be piezoelectrically excited and, thus, associated with a non-vanishing surface displacement profile. 
Lamb modes with angular frequency $\omega$ and radial wave vector $k_r$,  ${\bf u}(r, \theta, z)$  safisfying this constraint can be expressed as a superposition of a longitudinal (L) wave polarized along the propagation direction and a 
\cMM{MM: this polarization info is not very clear, maybe it can be left out}
 {transverse} (T) wave polarized in the $\hat r\hat z$ plane, both with wave vector component along the radial direction equal to  $k_r$. In the following, the acronyms $L$ and $T$ will be used to denote these components. The dispersion relation of these modes is given by the solutions of the following equation:

\begin{widetext}
\begin{equation}
  \left[ 
    \frac
      {\tan \left(\frac{d k_T}{2}\right) }
      {\tan \left(\frac{d k_L}{2}\right) }
  \right]^{i_s} 
   +
      \frac{4  k_L k_T k_r^2}
      {
        \left(k_T^2 - k_r^2\right) 
      \left[(v_L/v_T)^2  (k_L^2+k_r^2 ) - 2 k_r^2\right]} = 0,  
  \label{detSym}
\end{equation}

\noindent where $v_L$ and $v_T$ are, respectively, the longitudinal and tranverse velocities and $k_p^2 = \left( {\omega}/{v_p}\right)^2 - k_r^2$ ($p=L,T$) the wave vectors of their longitudinal and transverse components along $\hat z$. 
The exponent $i_s=\pm 1$ is related to the reflection symmetry of the mode envelopes relative to center plane $z=0$ of the disk: modes with $i_s= + 1$ are symmetric while those with $i_s= - 1$ are anti-symmetric (see Sec.~\ref{r-z-polarized modes} for details). 
The displacement field for the solutions with $i_s= 1$ is given by:

\begin{eqnarray}
  {\bf u}^n_{k_r}(r, \theta, z) &=&e^{i n \theta } B_e
  \left(
  \begin{array}{c}
    r^{-1} \left[ k_r r J_{n-1}(k_r r)-n J_n(k_r r)\right]  
    \left[ (k_T^2-k_r^2)\cos(k_L z) - 2 k_L k_T\cos( k_T z ) \right] \\
     - k_L J_n(k_r r)   \left[ \sin(k_L z) + 2 \frac{k_r^2}{(k_T^2-k^2_r)}\sin(k_T z) \right]  
  \end{array}
  \right)  
  \left( \begin{array}{c} {\hat r}\\ {\hat z}\\ \end{array} \right), 
  \label{uSym}
  \end{eqnarray}

\end{widetext}

\noindent where $B_e$ is the mode amplitude. The same expression applies for the $i_s=-1$ modes by interchanging $\sin\rightarrow \cos$  and $\cos\rightarrow \sin$. It is interesting to note that while the dispersion of these modes does not depend on the winding index $n$, their spatial profiles (or wave functions) do.

The black lines in Fig.~\ref{dispSymPlot}(b) show the  Lamb-wave dispersion for the elastically isotropic approximation for sapphire determined using Eq.~\ref{detSym}.  $k_r$ is normalized to  $k_{r,1}=\pi/\rfilm$, where $\rfilm=70~\mu$m is the radius of the piezoelectric film (this normalization will facilitate comparison with the Lamb modes of a disk with finite radius $\rfilm$ in Sec.~\ref{Radially confined Lamb modes}).  
Similarly to 1D Lamb modes (i.e., propagating along a single direction, cf.~Ref.~\onlinecite{Auld90a}), the dispersion for small $k_r$'s (i.e., $k_r\ll k_L, k_T$) can be approximated by the superposition of the quadratic dispersions of the  transverse and longitudinal wave components  given by: 

\begin{equation}
  \omega_{k_r} = v_p  \sqrt{k^2_p + k_r^2} \approx m_p v_p \frac{\pi}{d}\left[1+\frac{1}{2} \left(\frac{k_r}{k_p}\right)^2\right], 
  \label{EqOmega}
\end{equation}

\noindent where $p=L,T$ and $m_p=1,2,\dots$. 

The shape of the modes depend on the relationship between $k_r$ and the frequency $f$. For $k_r$ in the range  $[0,2\pi f/v_L]$, $k_L$ and $k_T$ are both real and the transverse and longitudinal mode \rAP{components}{componets} extend over the whole substrate. The dispersion branches can, however, interact and anti-cross, leading to a complex dispersion shape. Furthermore,  the solutions of  Eq.~\ref{EqOmega} fullfil the condition $k_p d/2=m_i$, for both $p=L,T$, i.e.,   both the transverse and longitudinal wave components have  an integer number of half cycles  within the substrate. For the present experimental situation,  $m_p>30$ is a large number. \cPVS{Check!}{Under these conditions, it can be easily shown that the frequency spacing between the Lamb modes around a given frequency becomes  
$\Delta f_\mathrm{Lamb} \approx  m_\mathrm{GCD} v_T/(2 d)$, where $m_\mathrm{GCD}$ is the greatest  common divisor\footnote{see, e.g., \href{https://en.wikipedia.org/wiki/Greatest_common_divisor}{https://en.wikipedia.org/wiki/Greates\_common\_divisor}} (or factor) of $m_L$ and $m_T$.
}

For increasing $k_r$, the mode frequencies asymptoptically approach first the linear dispersion of bulk longitudinal and then of the bulk transverse modes propagating along the radial direction, which are reproduced  by the blue and green dashed lines in Fig.~\ref{dispSymPlot}, respectively, with phase velocities $v_L$ and $v_T$. 
For frequencies $f$ satisfying  $2\pi f/v_L<k_r<2\pi f/v_T$, the wave vector \rAP{component}{componenet} $k_T$  remains real, but $k_L$ becomes imaginary. Their transverse components thus extend over the whole substrate, while the  longitudinal one  decays away from the  surfaces. The brown dashed line shows the dispersion of the so-called Lam{\'e} modes~\cite{Auld90a}, which are transverse vibrations  polarized along $\hat z$. Finally, for  $k_r>2\pi f/v_T$ all branches  converge to the single SAW branch with an acoustic displacement fully confined to the surface near region. The dispersion equation for cylindrical SAWs is  given by Eq.~\ref{EqSAW}.

 %
 %
 \begin{figure*}[tbhp]
  \centering
  \includegraphics[width=0.90\textwidth]{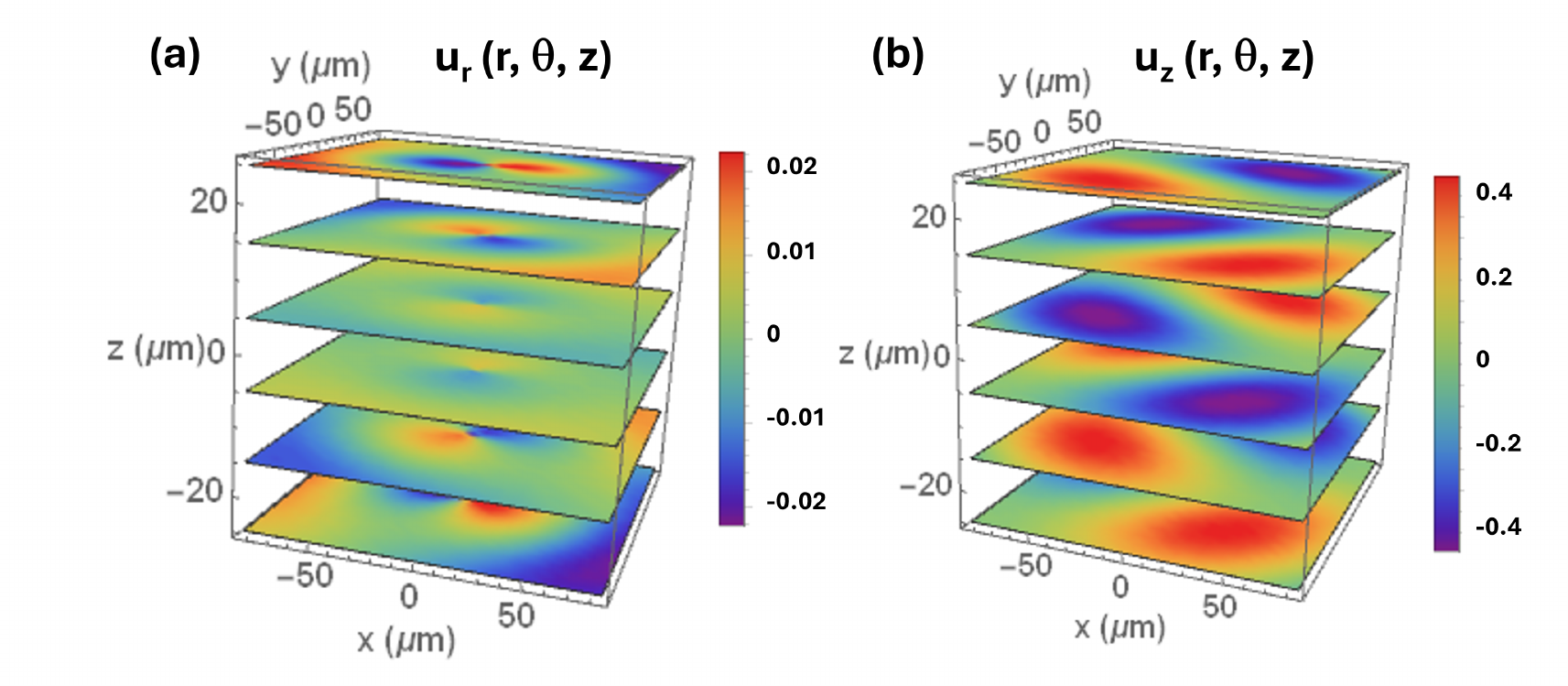}
  \caption{ 
    {\bf Long-wavelength helical Lamb wave profiles.}
    Displacement field along (a) $\hat r$ and (b) $\hat z$ calculated using Eq.~\ref{uSym} for a quasi-longitudinal mode with radial wave vector  $k_r=2\pi/\rfilm$ and winding number $n=1$. 
  } 
   \label{splicePlotL}
   \end{figure*}
  
The solutions of Eq.~\ref{EqOmega} for modes with longitudinal character with a small radial wave vector and  polarization close to $\hat z$  are of particular interest for us. These modes can be efficiently excited and detected by the BAWR since their strain field near the surface is almost colinear with the applied rf field, a configuration which yields the highest electromechanical coupling 
(cf.~Sec.~\ref{r-z-polarized modes}).
Figure~\ref{splicePlotL}(a) and \ref{splicePlotL}(b) display  maps for the radial and vertical components of the  displacement field for such a Lamb wave with a winding number $n=1$ and a small wave vector component (corresponding to $k_r=k_{r,1}=\pi/\rfilm$). Due to the small $k_r$,  the displacement amplitudes along $\hat z$ are much larger than the ones in the radial direction. The displacement field rotates counterclockwise both in time and as a function of depth creating a vortex at the center of the pattern.   

The rotation pattern at the sample surface of Fig.~\ref{splicePlotL} approximates  the interferometric maps for different phases in Fig.~\ref{Fig4}. Here, it is important to recall that the angular phase profile created by the three sector-BAWRs is not continuous but rather consists of three discrete steps [cf.~\ref{Fig4}(k)]. In order to reproduce the experimental configuration, one thus require a superposition of modes ${\bf u}^n_{k_r}(r, \theta, z)$  with different winding numbers $n$. 


\subsection{ZnO-induced surface modes}
\label{ZnO-induced surface modes}

The ZnO overlayer provides the piezoelectric properties required for electrical excitation as well as the thickness discontinuity for the radial confinement of the Lamb modes. The analysis of cylindrical Lamb modes in the previous section 
disregards this  layer, which is very thin compared to the substrate. This approximation is valid  for  Lamb modes with small $k_r$,  which extend over the whole substrate.
For large wave vectors $k_r\geq k_T$, in contrast, both $k_L$ and $k_T$ become imaginary (cf.~Eq.~\ref{detSym}) and the Lamb modes turn into SAWs with field fully confined near the surface: the ZnO overlayer can, in this case, significantly affect the acoustic propagation properties.  In particular, since the acoustic velocities of ZnO are significantly lower than the sapphire ones, the ZnO layer effectively creates a surface waveguide for acoustic waves that can support SAW-like modes with different transverse profiles.

To address the effects of the ZnO layer, we calculated the dispersion of SAW-like modes of a ZnO-coated sapphire substrate (see Sec.~\ref{SAW modes in ZnO-coated c-plane sapphire} for detail). The results are shown by the blue and purple dots in Fig.~\ref{dispSymPlot}(b). The calculations were performed for SAWs propagating along one dimension (i.e., the cylindrical character was not considered) and yield not only the frequency but also the piezoelectric generation efficiency (i.e., the electromechanical coupling efficienty $\keff$ of the modes.  Figure~\ref{dispSymPlot}(b) only includes modes with a moderate to high electromechanical coupling. SAW$_1$ and SAW$_2$ are  modes with a moderate electromechanical coupling. SAW$_3$ is a true SAW mode for large $k_r$. This mode becomes a pseudo-SAW  with a large electromechanical coupling ($\keff$ up to 6\%) when it enters the dispersion range of extended Lamb modes for small $k_r$'s. The short period modes detected in Figs.~\ref{Fig3} and \ref{Fig4} are attributed to this mode: indeed, the measured oscillation periods (red asteriscs in Fig.~\ref{dispSymPlot}) agree very well with the mode dispersion.  

\subsection{Radially confined Lamb modes}
\label{Radially confined Lamb modes}

We now address the radial confinement of the piezoelectrically excited modes, which results in the formation of drum-like Lamb modes underneath the ZnO film.





\begin{figure}[tbhp]
  \centering \includegraphics[width=0.50\textwidth]{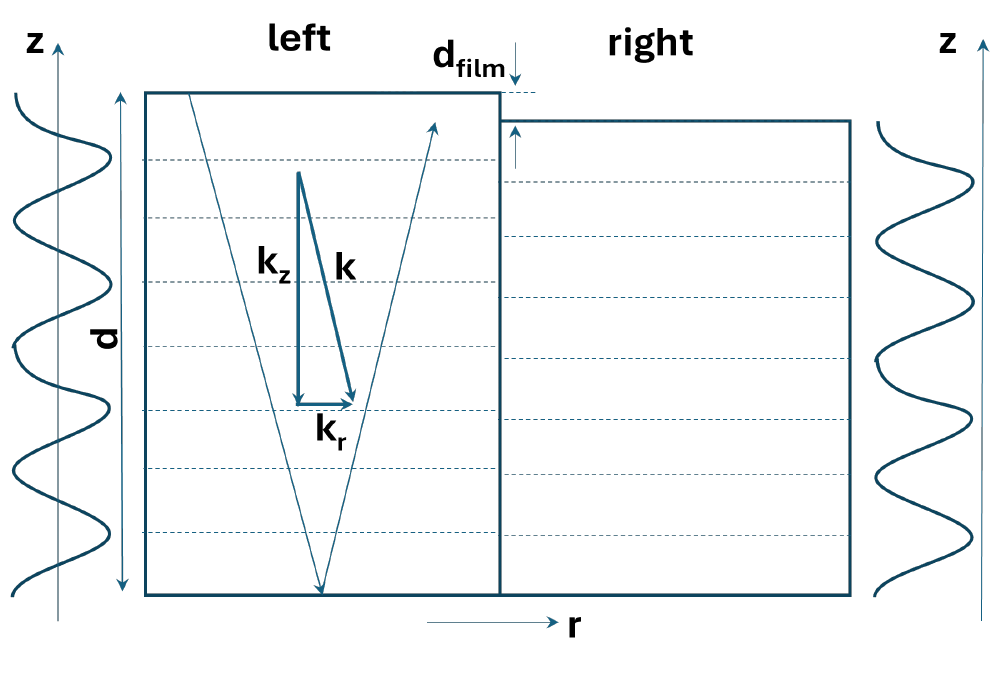}
\caption{ {\bf Radial confinement mechanism.} Propagation of Lamb waves with as small in-plane wave vector component $k_r$ across the interface between two plates with different thickness $\dfilm \ll d$ . The side plots displays the wave  profile at the left and right sides for frequencies $f_m$ and $f_m+\Delta f_m$, respectively, satisfying the boundary conditions at the top and back surfaces. The corresponding displacement anti-nodes are indicated by the dotted horizontal lines.
}
\label{FigSI-Impedance}
\end{figure}

The thickness discontinuity $\dfilm$ introduced by the ZnO layer creates Fabri-P{\'e}rot acoustic cavities with different thicknesses $d=d_\mathrm{sub} + \dfilm$ and  $d_\mathrm{sub}$ and  (cf.~Fig.~\ref{FigSI-Impedance}) as well as eigenmodes in the regions underneath and away from the ZnO film, which hinders  mode propagation between these two regions.
To  estimate the acoustic reflection coefficient at the lateral interface we assume that an essentially longitudinal Lamb eigenmode is excited on the left Fabri-P{\'e}rot cavity with frequency $f_m$ and wave vector $ {\bf k} = (k_r,  k_z) $ with a small radial component $k_r$. The in-plane velocity for that mode can be approximated by: 

\begin{equation}
v_r  \approx v_{L} k_r/|k_z|.  
\label{EqVr}
\end{equation}

The side panels of Fig.~\ref{FigSI-Impedance} display profiles for the wave displacement $u_z(r, z)$ eigenmodes of the left and right cavities with the smallest frequency difference $\Delta f_m$. These eigenmodes must satisfy the boundary conditions at the top and back surfaces for frequencies $f_m$ and $f_m+\Delta f_m$, respectively. These boundary conditions require  an integer number ($m_L$) of acoustic half cycles in each cavity, i.e., the mode frequencies are such that 
both the top and bottom surfaces are displacement antinodes (dotted horizontal lines in Fig.~\ref{FigSI-Impedance}). 

We will be here interested in the case where the overlayer thickness $\dfilm$ is much smaller than half of the wavelength along  $z$, i.e.,  $|\dfilm|<d/(2 m_L) = v_{L}/(4f_m)$. The eigenmodes with closest frequencies in the two cavities will then have wave vectors with z-components given by  $ (m_L-1) \pi /d$ and $ m_L \pi /(d - \dfilm)$ in the left  and right  plates, respectively. By using Eq.~ \ref{EqVr}, it is then straightforward to show that the relative mismatch in frequency ($\Delta f_m/f_m$) or radial velocity  ($\Delta v_r/v_r$)  along the in-plane direction becomes:

\begin{equation}
r_f = \frac{\Delta f_m}{f_m} =\frac{\Delta v_r}{v_r} = -\frac{\dfilm}{2 d}.
\label{Eqra}
\end{equation}

\noindent This approximation applies for $ \dfilm < \frac{d}{m_L} = \frac{v_{L}}{4f_m}$.
For a given $\dfilm$, the maximum frequency mismatch and reflection coefficient is inversely proportional to the frequency, and, for a fixed frequency, reduces with increasing substrate thickness.  

Using the nominal ZnO thickness used in our samples, one calculates a ratio $r_f = \Delta f_m/f_m \approx 0.08\% = 1/1250$. This value exceeds the measured inverse quality factor  $1/Q=1/3800$ of the extended Lamb modes by approximately a factor of 3. The high acoustic quality factor cavities reduces the overlap between  the frequency spectra at the two sides of the lateral interface, leading to an efficient acoustic reflection at the ZnO boundaries [cf.~Figs.~\ref{Fig1}(c) and \ref{Fig1}(e)]. High Q's are thus essential for an efficient confinement yielding  drum-like Lamb modes.
Finally, while derived for extended Lamb waves with a small wave vector ratio $k_r / k_L$, Eq.~\ref{FigSI-Impedance} also applies for SAWs provided that one replaces the  thickness $d$ by the penetration depth $\approx 1/\lSAW$ of the acoustic field, where $\lSAW$ is the SAW wavelength. This yields significantly larger reflections for SAWs as for extended modes.



In the absence of radial confinement, there is no restriction on the radial wave vector $k_r$ and the Lamb modes have a continuous frequency spectrum [cf. Fig.~\ref{dispSymPlot}(b)]. The situation changes if the modes become confined underneath the ZnO film: the available values for $k_r$ will then depend on the radial boundary conditions at the boundaries $r=\rfilm$ of the confinement area. A closed analytical solution for the confined field can, in general, not be derived. A very good approximation for small $k_r$ (i.e., $k_r\ll k_T, k_L$) is obtained by imposing  at $r=\rfilm$ the constraint:

\begin{eqnarray} 
J_n(k_r \rfilm)=0 &\rightarrow& \nonumber \\
k_{r,m_r} &=& \frac{1}{\rfilm}Z_{b_n}(m_r) \approx \sqrt{m_r^2 + n^2} \frac{\pi}{\rfilm}\nonumber,
\label{EqZB}
\end{eqnarray}

\noindent where $m_r = k_r \rfilm/\pi$ and $Z_{b_n}(m_r)$ is the zero of $J_n$ of order  $m_r$. The approximation on the right hand side is justified in Sec.~\ref{Solution for a disk of radius a}. For $n=0$, one retrieves Eq.~\ref{Eqkr}: the confined solutions are then the dispersion points in Fig.~\ref{dispSymPlot} for which $m_r$ is an integer, i.e., $m_r=1,2,\dots$.   Interestingly, this approximation applies for both a stress-free as well as for a clamped lateral surface  at  $r=\rfilm$. 

The red dashed lines superimposed on the simulation results of Figs.~\ref{Fig1}(e) and \ref{Fig1}(f) display profiles for the $u_z(r)   \propto J_0(k_{r, m_r})$ (cf. Eq.~\ref{uSym}) calculated using $k_{r,m_r}$ values determined using  Eq.~\ref{EqZB} for $m_r=1$ and $m_r=2$, respectively. The Bessel function approximation reproduces well the finite element field results.

We now proceed to the determination of the frequency shifts induced by the radial confinement.
By combining Eqs.~\ref{EqOmega} and \ref{EqZB}, we obtain the following expression for the frequency of the drum modes of small $k_r$ with longitudinal and radial indices $m_L$ and $m_r$, respectively: 
\begin{equation}
%
  f(m_L, m_r) = f_{m_L} +   \frac{v^2_L Z_{b_n}^2(m_r) }{8 f_{m_L} \pi^2 \rfilm^2}, f_{m_L} = m_L v_L \frac{\pi}{d}.
  \label{Eqfmr}
  \end{equation}

\noindent The radial drum modes thus add a series of resonances to the longitudinal substrate modes (index $m_L$) with frequency spacing dictated by the radius of the ZnO film.

We attribute the extra resonances observed  in the electrical response [cf.~Figs.~\ref{Fig1}(b) and \ref{Fig2}(b)] close to main longitudinal modes to the excitation of the radial drum modes given by Eq.~\ref{Eqfmr}. 
In fact, the splitting $\Delta f_1=7$~MHz between the first two modes in Fig.~\ref{Fig1}(b) is very close to the splitting of 7.7~MHz calculated from Eq.~\ref{Eqfmr} for $f_{m_L} = 1.04$~GHz. Furthermore, the splitting $\Delta f_1=2.5$~MHz in Fig.~\ref{Fig2}(b)  agrees very well with the predicted value of 2.7~MHz for $f_m=3$~GHz. 

The prediction for the  splitting between the third and the first lines of 7.5 MHz, however, exceeds the measured value of 4.4~MHz. The discrepancy may be associated with an increased interaction between the  dispersion branches in Fig.~\ref{dispSymPlot}, which is not accounted by the simple model of Eq.~\ref{Eqfmr}. Further work will be required to explain these deviations.

The observation of the resonances in the electrical reflection together with the quantitative agreement with the  model thus provide a further unambiguously  evidence for radial confinement of  the small $k_r$  modes. Here, we remind ourselves that, in contrast to the radial confinement of short period vibrations established by Fig.~\ref{Fig3}(d), the radial confinement of long-period vibration could not be established by interferometric \rAP{measurement}{measumrents} due to their small surface displacements.

\section{Conclusions}
\label{Conclusions}

We have demonstrated a novel concept for the piezoelectric generation of drum-like helical acoustic Lamb waves in the GHz range. The concept  exploits the vertical confinement  via acoustic reflections at the back side of the substrate as well as the radial confinement by substrate thickness variations. We have shown here that the concept can be easily extended for the generation of helical  Lamb modes, which, via  acousto-optical interactions, produce light beams with tunable orbital angular momentum.

The Lamb-wave approach enables reaching considerably higher frequencies than those so far reported for helical beam generation  using  arrays of interdigital SAW transducers (IDTs)~\cite{Riaud_PRA4_34004_15} or IDTs with ring~\cite{Laude_APL92_08} or spiral-like shapes~\cite{Riaud_PRA7_24007_17}. On the one hand, these SAW-based processes can also profit from the lateral confinement using overlayers introduced here. 
On the other hand, the Lamb-wave approach can be straight-forwardly extended to the spiral-shaped BAWR geometries. 

While for the spiral-type geometries the  helicity sign is fixed by the structure shape,  approaches using arrays of SAW's or BAWR's enable dynamic control both the magnitude and polarity of the helicity, thus offering extra flexibitity for signal processing. Due to the finite number of transducers, these approaches, however, normally generate a superposition of modes with different winding numbers $n$.  One of the challenges  is then to increase the number of phase-synchronized transducers for the generation of modes with a well-defined $n$.

The Lamb-generation and radial confinement rely of the formation of a high-Q acoustic cavity in the substrate: they thus require substrates with low acoustic absorption. The radial confinement can be substantially enhanced by increasing the thickness contrast ratio between the confinement region and the surroundings (cf. Eq.~\ref{Eqra}), e.g., by using thicker overlayers or reducing the substrate thickness.  Alternatively, the acoustic field can be vertically confined by using underetching or via the use of distributed Bragg reflectors (DBRs), which have become conventional processes for the BAWR technology.  

We now briefly address prospects for the enhancement of the acousto-optical generation efficiency of helical optical beam. 
Due to the small displacement amplitudes ($\ll\lambda_\mathrm{opt}$), the phase modulation $\delta \phi(r,\theta)$ imprinted via reflection is very small.  Large modulation amplitudes can be achieved by Bragg back-scattering of an incoming  light beam in the standing acoustic mode formed in the substrate. 
Phase matching for the acousto-optical interaction requires, in this case, an optical wavelength $ \lambda_\mathrm{opt} = 2 n_{ref} v_{LA} /(m\fBAW$), where $n_{ref}$ is the substrate refractive index and $m=1, 3, \dots$ the diffraction order. 
For the optical wavelength $\lambda_\mathrm{opt}=532$~nm employed in the present experiments, this condition can only be fulfilled for very high acoustic frequencies (i.e., approx 74~GHz and 25~GHz for first and third-order diffractions, respectively). For an optical wavelength in the near-infrared $\lambda_\mathrm{opt}=1550$~nm, however, the third-order Bragg condition is satisfied at a much lower frequency $\fBAW=8.5$~GHz.
%
%
%
%

Finally a particularly interesting approach exploits the simultaneous confinement of light and vibrations formed  in semiconductor  microcavities~\cite{Trigo02a,Lacharmoise04a,PVS312,PVS318}. 
The strong overlap between the optical and vibrational fields ensures a large acousto-optical coupling, which can be further enhanced by introducing quantum well excitonic resonances to form  light emitters of $\mu$m sized based on exciton polaritons and exciton polariton condensates~\cite{PVS334}.  Modulation of these light emitters  by piezoelectrically stimulated bulk acoustic vibrations has been demonstrated~\cite{PVS334}: the helical Lamb-wave concepts  introduced here can  impart OAM to the emission from confined polaritons and their Bose-Einstein-like condensates. The latter provide a powerful approach for the efficient generation of chiral optical beams with tunable OAM. In addition, it enables exploitation of the rich physics of \rAP{rotating}{stirred} polariton  condensates under an acoustically induced stirring, which has so far only been addressed using optical stirring methods\cite{Redondo_NL23_4564_23,Gnusov_SA9_23,arXiv_2305_03782v1}. 

\section*{Methods}
\label{Experimental details}

\subsection*{Sample structure and fabrication}
 The devices [cf.~Fig.~\ref{Fig2}(a)] were fabricated starting  from the bottom BAWR metal contact, a 10~nm Ti/30~nm Au layer stack deposited on a 450~$\mu$m-thick c-plane sapphire substrate. The bottom BAWR contact was then coated with a $\dfilm = 700$~nm-thick layer of textured piezoelectric ZnO film capped with a 20-nm-thick protective SiO$_2$ layer using rf-magnetron sputtering at 150$^\circ$. This ZnO thickness yields a center resonance frequency for longitudinal acoustic (LA) modes of approx $v_{L}/(2 \dfilm)\sim 4$~GHz. The BAWR are, however, able to excite vibrations over a wide range of frequencies extending down to less than 1~GHz. The ZnO layer was then subsequently etched to form a circular disk with radius  $\rfilm$. 
 In the final step, the top contact, a 10~nm~Ti/50~nm~Al/10~nm~Ti layer stack was fabricated using a photolithographic lift-off process. 
 BAWRs with different dimensions were fabricated and tested: we will report here only on a device with the geometric parameters given in the caption of Fig.~\ref{Fig1}.
 
 \subsection*{Numeric simulations}
 {Finite-element simulations of the sample structure were performed using COMSOL Multiphysics by assuming  the ZnO film and the sapphire substrate to be  isotropic, i.e., ignoring the in-plane acoustic anisotropy of the trigonal sapphire substrate (cf. Sec.~\ref{appendix:Acoustic waves in ZnO-coated sapphire}). Under this assumption, the acoustic displacement field ${\bf u}$ can be conveniently determined in cylindrical coordinates using a two-dimensional model. 
 }
 
 \subsection*{Acousto-electrical and -optical responses}

 The rf response of the BAWRs was investigated using a vector network analyzer (VNA) to measure the electrical scattering parameters $s_{ij}$ [cf.~Fig.~\ref{Fig2}]. The sector-like BAWRs [cf.~Fig.~\ref{Fig2}(a)] are excited by independent radio-frequency (rf) drives. This configuration enables probing each BAWR individually, which is advantageous for the determination of the acoustic coupling between them, as well as for the excitation of helical waves. For the latter, the BAWRs were driven by with rf phases differing by 120 degrees between neighboring sectors, opportunely controlled using an rf phase shifter (cf. Sec.~\ref{Radio-Frequency phase shifter}).\\ 
 %
 Maps of the vertical surface displacement $\delta z$ were obtained using a scanning Michelson interferometer with a spatial resolution of approximately $1~\mu$m  [cf.~Fig.~\ref{Fig2}(a)]. In order to provide detailed information about the relative phases of the displacement, the interferometric signal from a fast photodetector (bandwidth of 2~GHz) is connected to the VNA delivering the  rf drive. Further details about the phase sensitive detection can be found in the supplement of Ref.~\onlinecite{PVS367}.

\section*{Acknowledgments} 
We acknowledge the funding from German DFG (grants 359162958 and 426728819). AP acknowledges funding from the Alexander von Humboldt foundation through the Experienced Researcher fellowship programme.
The authors thank A. S. Kuznetsov for discussions and for a critical review of the manuscript as well as the technical support by S. Rauwerdink, W. Anders, N. Volkmer, and A. Tahraoui. 







\def\litdir{c:/myfiles/jabref}
\def\litdir{x:/sawoptik_databases/jabref}    


%


\clearpage
\newpage
\newpage
\widetext

\beginsupplement





\begin{center}
{\bf SUPPLEMENTARY INFORMATION:\\
GHz helical acoustic drum modes on a chip\\
}

{
	N. Ashurbekov$^1$, I. dePedro-Embid$^1$, A. Pitanti$^{1,2}$, M. Msall$^{1,3}$,  
    and P. V. Santos$^1$\\
\it $^1${Paul-Drude-Institut für Festkörperelektronik, Leibniz-Institut im Forschungsverbund Berlin e. V., Hausvogteiplatz 5-7, 10117 Berlin, Germany}\\
\it $^2${University of Pisa, Dipartimento di Fisica E. Fermi,
largo Bruno Pontecorvo 3, Pisa 56127, Italy}\\
\it $^3${Department of Physics and Astronomy, Bowdoin College, Brunswick, Maine 04011, USA}
}
\end{center}

We present in this document additional material to support and complement the conclusions of the main manuscript.

\section{Acoustic waves in ZnO-coated sapphire}
\label{appendix:Acoustic waves in ZnO-coated sapphire}

The c-axis oriented sapphire substrate belongs to the trigonal group. The solution of the wave equation  (see  Eq.~\ref{appendix:pzEq1}, see below) for this material for wave propagating with wave vector $\bf q$ in the a-b crystallographic plane (i.e., in the radial ($\bf r$) direction  perpendicularly to the crystallographic c-axis) yields three types of acoustic waves:\cite{Auld90a}  
 \begin{itemize}
 \item a pure longitudinal wave polarized along the propagation direction ($\bf r$) and phase velocity $v_{L}$; 
 \item a pure shear wave polarized along z and phase velocity ($v^{-1}_{T,z}$);  
 \item a pure shear wave polarized along z and phase velocity  ($v^{-1}_{T,\theta}$)
 
 \end{itemize}  

 \noindent In addition to these bulk modes, the substrate also support surface acoustic modes (SAWs) confined to the surface region.

\begin{figure}[bthp]
	\centering
		\includegraphics[width=0.6\textwidth]{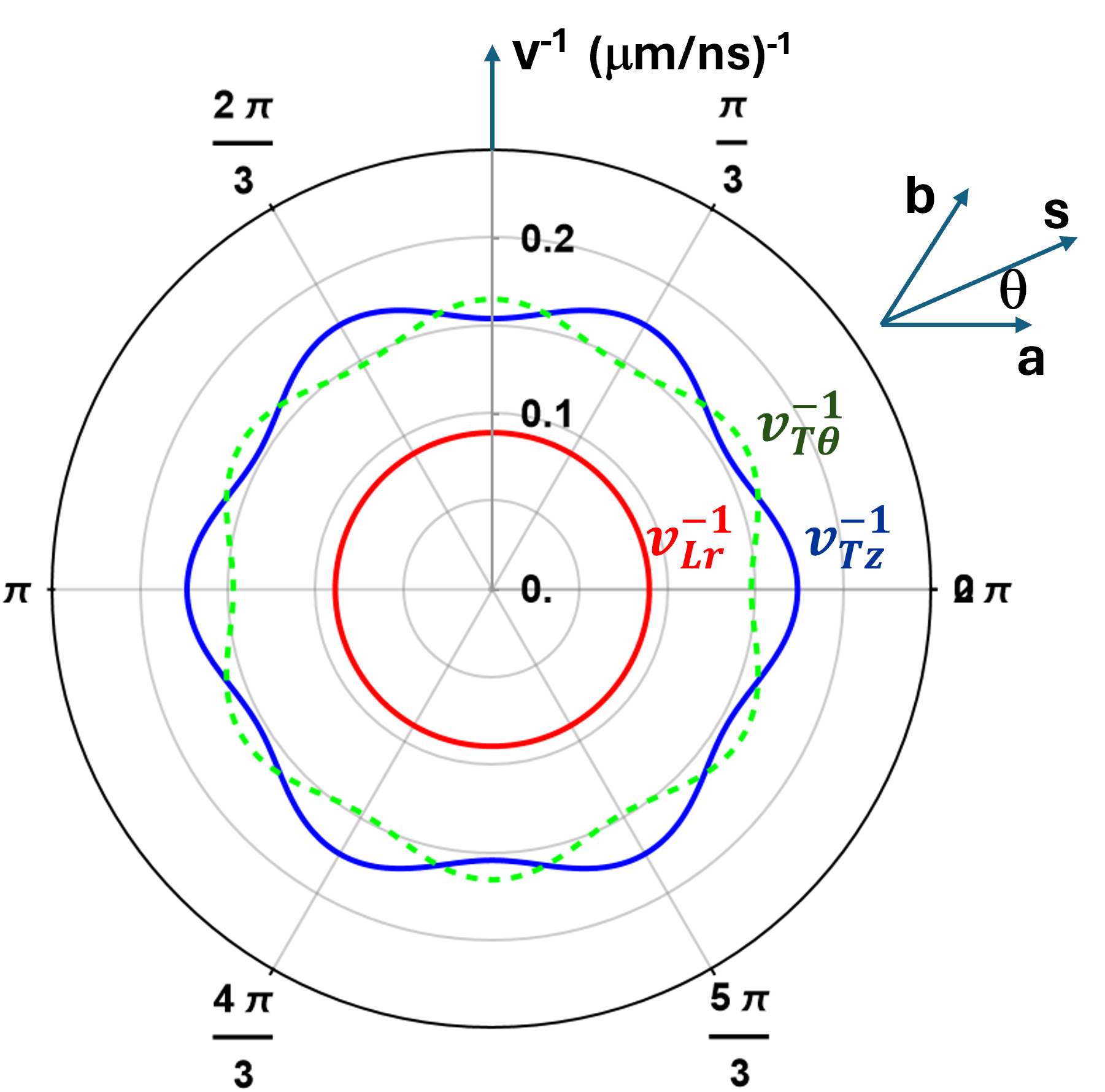}
\caption{
Angular dependence of the acoustic slowness for acoustic waves propagating with a wave vector $\bf k_r$  along the c plane  of sapphire for longitudinal ($v^{-1}_{Lr}$) and transverse modes polarized along ($v^{-1}_{Tz}$) and perpendicularly ($v^{-1}_{T\theta}$) to the c-axis. $a$ and $b$ are the in-plane crystallographic axes.  }
\label{Slowness_Sapphire}
\end{figure}

The phase velocities of these modes depend on their propagation angle with respect to the $a$ crystallographic axis. The slowness surfaces ($v_{i}^{-1},$ i.e., the inverse velocity surfaces) for these waves are illustrated in Fig.~\ref{Slowness_Sapphire}. 
The elasticity constants for crystalline sapphire used for  the calculations in Fig.~\ref{Slowness_Sapphire} (as well as for crystalline ZnO and for the isotropic approximation for sapphire)  are listed in Table~\ref{appendix:pTable}.  Table~\ref{appendix:vTable} lists representative bulk and surface wave speeds for sapphire. While the phase velocity of the longitudinal mode is isotropic in the plane, the one for the transverse modes varies with the azymuthal angle. This angular variation in phase speed introduces anisotropies in energy flux, including concentration of the acoustic energy along caustics determined by inflections in the constant frequency surfaces in k-space, an effect known as \cMM{DONE-ref J.P. Wolfe book?}{phonon focusing}~\cite{Wolfe_Book_98}.  

 Finally, the c-plane sapphire substrate supports a single surface acoustic (SAW) mode with the phase velocities $v_\mathrm{SAW}$  along the $a$ and $b$ axes listed in the last column of the table.

\section{Acoustic modes of a cylindrical plane}
\label{appendix:SI_Acoustic Modes}

A very instructive description of Lamb waves propagating along a fixed direction using cartesian coordinates can be found  in Chapter 10 of Auld's classic book on acoustics [~\onlinecite{Auld90a}]. Here, we extend Auld's one-dimensional formulation  for the acoustic displacement field $\vec u$ of cylindrical Lamb waves propagating  in a disk of thickness $d_\mathrm{sub}$, as displayed schematically in Fig.~\ref{Fig1}(a) of the main text.  

We will assume the disk material to be elastically isotropic with non-vanishing elastic constants $c_{11}$ and $c_{12}=c_{13}$. For numerical calculations, we will use the parameters  listed in Table~\ref{appendix:vTable} [row isotropic "sapphire (iso)"]. 
This approximation, on the one hand, disregards the presence of the ZnO film as well as the (quite pronounced) azymuthal dependence of the elastic properties of c-plane sapphire on  the propagation angle $\theta$  (cf. Fig.~\ref{Fig1}(a), main text) in the a-b surface plane.  On the other hand, it considerably simplifies the problem, enabling the derivation of analytical expressions for the field distribution. The impact of the ZnO layer will be considered as a perturbation in \cMM{MM: somehow not referencing right section} {Sec.~\ref{ZnO-induced surface modes}} of the main text. Finally, we will only consider modes with a real propagation constant along the radial direction.

\begin{table}[btp]
  \caption{Densities ($\rho$, in units of $kg/m^3$) and non-vanishing elastic constants  ($c_{ij}$, in units of $10^{12}N/m^2$) for the materials used in this work.}
  \label{appendix:pTable}
  %
  \begin{tabular}{ |l|p{0.9cm}|p{0.9cm}|p{0.9cm}| p{0.9cm} | p{0.9cm} | p{1.1cm} | p{0.9cm} | p{0.9cm} | p{1.1cm} | p{0.9cm} |} 
      \hline
  \hline
  Material        & $\rho$      	&$c_{11}$	& $c_{12}$& $c_{13}$& $c_{33}$ & $c_{14}$ & $c_{44}$ & $c_{55}$ & $c_{56}$ & $c_{66}$\\
  \hline                   
  Sapphire\footnote{https://www.crystran.com/optical-materials/sapphire-al2o3}	  
  &3986 		&  0.497 	& 0.163 	& 0.111 	& 0.498 	& -0.024 	&  0.147 	&  0.148 	& -0.024 	& 0.167\\
  Sapphire (iso)\footnote{From COMSOL Multiphysics.}	  &3980 	&  0.538 	& 0.231 	& 0.231 	& 0.538 	& - 		&  0.154 	& 0.154 	&  -  	& 0.154\\
  ZnO\cite{LB11}	         &5665 		&  0.2096 	& 0.120 & 0.1046	& 0.211 	& - 		&  0.0423 	&  - 		&  - 		& 0.444\\
  \hline
  \hline
  \end{tabular}
  \end{table}

\begin{table}
\caption{Calculated longitudinal  ($v_{L}$) and transverse ($v_{T,i}$) acoustic velocities for modes with polarization along $i$ propagating along the $a$ and $b$ axes of a c-plane sapphire substrate. 
The last column summarizes the corresponding  velocities of the SAW modes 
on a plain sapphire substrate. 
The calculations were carried out following the procedure delineated in Refs.~\onlinecite{Auld90a} and \onlinecite{PVS156} using the parameters for sapphire and ZnO listed in Table~\ref{appendix:pTable}.
  }
\label{appendix:vTable}
\begin{tabular}{ | l | c | c | c | c | } 
\hline
\hline
Material        & $v_{L}$& $v_{T,\theta}$& $v_{T,z}$& $v_{SAW}$  \\  

                & (m/s)     & (m/s)     & (m/s)     & (m/s)    \\
\hline
Sapphire a-axis & 11170     & 6767      & 5742.88   & 5702     \\
Sapphire b-axis &  11184    & 6040      & 6473      &  5556    \\
Sapphire iso    & 11622.7   & 6212.62   & 6212.62   & 5761.66  \\
\hline
\hline
\end{tabular}
\end{table}



The dynamical equation for elastic deformation $\vec u$ can be expressed as:~\cite{Auld90a}

\begin{equation}
\rho\frac{\partial^2 \Vec{u}}{\partial t^2} - \nabla\cdot {\bf \tau} = \Vec{f},
\label{appendix:pzEq1}
\end{equation}

\noindent where $\rho$ is density, $\Vec{u}$ is the displacement field, $\Vec{f}$ the volume force, and $\tau$ the stress tensor. For the calculation of the eigenmodes, we will assume here that $\Vec{f}=0$. The wave equation is complemented by the following constitutive relations for a non-piezoelectric material\cite{Lerch_ITUFFC37_233_90}

\begin{equation}
\tau = c \varepsilon = c \nabla_s u 
\label{piezoEq9b}
\end{equation}

\noindent where $\varepsilon$ is the strain tensor, $c$ the elastic tensor, and $\nabla_s=\frac{1}{2}(\nabla + \nabla^T)$, where $\nabla$ is the gradient operator. In the Vogt notation, the strain vector becomes: $\varepsilon = [s_1, s_2, s_3, s_4 (yz), s_4 (xz), s_5 (xy)]^T$. 



If the propagation medium  is isotropic, these modes do not depend on the propagation angle $\theta$. When expressed in cylindrical coordinates, the problem stated by Eq.~\ref{appendix:pzEq1} becomes 2D-dimensional in the variable space $\{r, \theta, z\}$  with the following types of solutions:

\begin{itemize}
\item a pure longitudinal wave polarized along the propagation direction ($\bf r$) and phase velocity $v_{Lr}$; 
\item a pure shear wave polarized along z and phase velocity ($v_{Tz}$);  
\item a pure shear wave polarized along z and phase velocity  ($v_{T\theta}$)
\end{itemize}  

Different approaches for the solution of Eq.~\ref{appendix:pzEq1} are reviewed by Honarvar et al. in Ref.~\onlinecite{Honarvar_IJSS44_5236_07}. In the most general case, the acoustic field $\vec u(r, \theta, z, t)$ can be decomposed into three components~\cite{Honarvar_IJSS44_5236_07} expressed in terms of scalar potential functions $\phi(r, \theta, z, t)$, $\chi(r, \theta, z, t)$, and $\psi(r, \theta, z, t)$. The component associated with $\phi(r, \theta, z, t)$  has compressional character, the other two are dilation-free (i. e., equivoluminal) shear components polarized along $z$ [$\xi(r, \theta, z, t)$] and in the plate disk plane (i.e., perpendicular to $z$) [$\xi(r, \theta, z, t)$]. Following Honarvar {\it et al.},  the displacement field $\vec u(r, \theta, z,t)$ can be expressed as:

\begin{equation}
	\vec u(r, \theta, z,t) = \nabla\phi(r, \theta, z,t) + \nabla \times (\chi(r, \theta, z,t)\hat e_z )+ a\nabla\time \nabla \times (\psi(r, \theta, z,t)\hat e_z ),
		\label{appendix:Equ}
\end{equation}

We search for solutions of Eq.~\ref{appendix:Equ} consisting of a superposition of  Bessel functions  of the first kind of order $n$, $J_n(k_r r)$, where $n=0,1,2,\dots$ is the winding number. The radial wave vector  $k_r$ yields the spatial periodicity of the oscillation along $\hat r$ for large radial $r$ (i.e., in the limit where the Bessel functions become essentially a cos-like function of $r$).  We search for solutions based on trial functions of the following form:

\begin{eqnarray}
\phi(r, \theta, z, t) &=& (B^+_n e^{\imath k_z z } 
) e^{\imath ( -\omega t + n  \theta})  J_n( k_r r)  \label{appendix:Eqphi} \\
\chi(r, \theta, z, t)  &=& (C^+_n e^{\imath k_z z } 
) e^{\imath ( -\omega t + n  \theta})  J_n( k_r r)  \label{appendix:Eqchi} \\
\psi(r, \theta, z, t) &=& (D^+_n e^{\imath k_z z } 
) e^{\imath ( -\omega t + n  \theta})  J_n( k_r r).  \label{appendix:Eqpsi} 
\end{eqnarray}

\noindent Here, $k_z$ is a wave vector component along z. The solution with $n=0$ is the fundamental mode while the ones with $n>0$ are helical modes with a vortex at $r=0$.  The winding number $n$ describes the number of windings of the acoustic field during a full rotation around the $z$ axis (i.e., as the azimuth angle $\theta$ varies from $0$ to $2\pi$), corresponding to the topological charge of the vortex.

By substituting the  previous expressions into Eq.~\ref{appendix:Equ}, one obtains the following expression for the displacement field $\vec u$  in the vector basis $\{B^+_e, C^+_e, D^+_e\}$ as

\begin{eqnarray}
&{\vec u}&(r,\theta,z,t)=\frac{1}{r} e^{i ( k_z z + n\theta)} \times \label{appendix:UdispMat}\\ 
&&
\left(
\begin{array}{ccc}
 k_r r J_{n-1}(k_r r)-n J_n(k_r r) & i n J_n(k_r r) & i a k_z (k_r r J_{n-1}(k_r r)-n J_n(k_r r)) \\
 i n J_n(k_r r) & n J_n(k_r r)-k_r r J_{n-1}(k_r r) & -a k_z n J_n(k_r r) \\
 i k_z r J_n(k_r r) & 0 & a k_r r^2 J_n(k_r r)  \label{uFull}\\
\end{array}
\right)
\left( \begin{array}{c} B^+_e\\C^+_e\\D^+_e \end{array} \right) \nonumber
\end{eqnarray}

With this expression for ${\vec u}(r,\theta,z,t)$, the wave equation (Eq.~\ref{appendix:pzEq1}) can be cast as an eigenvalue problem as:

\begin{equation}
(M_{u} - \rho \omega^2 I_3 )
\left( \begin{array}{c} B^+_e\\C^+_e\\D^+_e \end{array} \right)  = 0
\label{appendix:eigEqMat}
\end{equation}

\noindent where $I_3$ is the $3\times 3$ identity matrix and 

{\small
\begin{equation}
M_u = \left(
\begin{array}{ccc}
  \rho \omega^2_L&0&0\\
  0& \rho \omega^2_{T\theta}& 0\\
  0&0& \rho \omega^2_{Txz}\\
\end{array}
\right). 
\label{appendix:eigEqMatT}
\end{equation}
}

As expected for wave propagation in an isotropic medium, the eigenvectors of $M_u$ consist of a pure longitudinal wave polarized along the propagation direction with eigenvalue given by 
$ \omega^2_L =  v^2_L (k^2_z + k_r^2)$, with $v^2_L=c_{11}/\rho$,
and two degenerate transverse modes with eigenvalues 
$\rho \omega^2_{T\theta}=\rho\omega^2_{Txz} = v^2_T k_T^2 = v^2_L(k^2_z + k_r^2)$, with $v^2_T=c_{44}/\rho$.  $v_L$ and $v_T$ are, respectively, the longitudinal and transverse propagation velocities. The transverse modes have polarization in the $\theta$ and $xy$-planes, respectively.
\noindent  It is interesting to note that $M_u$  as well as its eigenvalues $\rho \omega_i^2 (k_r, k_z)$ do not depend on the winding number $n$, while the eigenvectors do.
The corresponding eigenvectors are:

  
  \begin{eqnarray}
    {\bf \hat u}_L(r,\theta ,z) &=&  \frac{1}{r}e^{i(k_L z +n\theta)}
  \left[  
    +  \left(r k_r J_{n-1}\left(r k_r\right)-n J_n\left(r k_r\right)\right){\bf \hat r}
    i n  J_n\left(r k_r\right) {\bf \hat \theta} 
    + i r  k_L J_n\left(r k_r\right){\bf \hat z} + 
    \right] \label{Eq_uL}\\
  {\bf \hat u}_{T\theta}(r,\theta ,z)  &=& \frac{1}{r}e^{i(k_T z +n\theta)}
  \left[ 
    + i n  J_n\left(r k_r\right){\bf \hat r}
    +\left(n J_n\left(r k_r\right)-r k_r J_{n-1}\left(r k_r\right)\right){\bf \hat \theta}
     \right] \label{Eq_uTt} \\
  {\bf \hat u}_{Txz}(r,\theta ,z)  &=& \frac{1}{r}e^{i(k_T z +n\theta)}
  \left[
     +i a  k_T \left(r k_r J_{n-1}\left(r k_r\right)-n J_n\left(r k_r\right)\right){\bf \hat r}
     -a n  k_T J_n\left(r k_r\right){\bf \hat \theta}   + a r  k_r^2 J_n\left(r k_r\right){\bf \hat z} 
     \right] \label{Eq_uTxy}.
  \end{eqnarray}
  
  

  

\subsection{Solution for a disk of  infinity  radius}
\label{Solution for a disk of  infinity  radius}


The eigenstates in Eqs.~\ref{Eq_uL}-\ref{Eq_uTxy} are the solutions of the wave equation for modes with a given frequency $\omega$, winding number $n$,  and (real) radial wave vector $k_r$ propagating along the axis of a cylinder with infinity radius and length. Solutions for modes propagating with a given wave vector along the axis of an infinity long  cylinder with a given radius are discussed  in Ref.~\onlinecite{Honarvar_IJSS44_5236_07}. We will consider in this section wave propagation in a disk of infinite radius but now delimited by free surfaces at $z=\pm d/2$. 

In order to fullfill the boundary conditions at the disk surfaces, we will choose a basis (cf.~Eqs.~\ref{Eq_uL}-\ref{Eq_uTxy}) consisting of modes with a given wave vector components $k_r$ along the radial direction. The solutions can be expressed as superpositions of the eigenstates expressed by Eqs.~\ref{Eq_uL}-\ref{Eq_uTxy} with counter-propagating wave vectors $\pm k_z$ along the $z$ direction fullfilling the boundary conditions at the disk surfaces. The corresponding mode coefficients will be denoted as $\left\{ B^\pm_e, C^\pm_e, D^\pm_e \right\}$.  
The boundary conditions at the free surfaces require vanishing stress ($\tau$) components perpendicular to the boundary surfaces $z=\pm d/2$, i. e., $\tau_{zz}=\tau_{rz}=\tau_{\theta z}$.    When these constraints are applied, the solutions fall into two types of modes: a pure shear wave and a wave with mixed longitudinal and shear displacements.

\subsubsection{Pure shear modes}
\label{theta-polarized modes}

Purely transverse modes are associated with the scalar potential $\psi(r, \theta, z, t)$. They are  polarized in the  $r-\theta$ plane and propagate in the z-direction.  These modes are decoupled from the modes with $r-z$ polarization (see below). Their displacement field  can be written as:

\begin{equation}
{\bf u}(r, \theta, z) =i C_e \frac{ e^{i \theta  n}}{r} \cos \left(  k_T z \right) 
 \left(
\begin{array}{c}
 -i n J_n\left(r k_r\right)  \\
 r J_{n-1}\left(r k_r\right) k_r-n J_n\left(r k_r\right) 
\end{array}
\right)
 \left(
 \begin{array}{c}
 \hat r\\
\hat \theta
\end{array}
 \right),
\label{eqPureShear}
\end{equation}

\noindent where $C_e$ is the mode amplitude. For $n=0$ these modes become a purely transverse mode polarized perpendicularly to the $z$ axis.  The wave vector of the modes along $z$,  $k_T$, must be related to the mode angular frequency by the boundary conditions at the top and bottom surfaces of the plate according to:
\begin{equation}
  k_z^2 = k_T^2 = \left( \frac{\omega}{v_T}\right)^2 - k_r^2 = m_T \frac{2\pi}{d}, \quad m_T=1,2,\dots
\label{eqkT}
\end{equation}

\noindent Here,  $v_T=\sqrt{c_{44}/\rho}$ is the transverse acoustic velocity and the coefficient $C_e$  the wave amplitude.   These modes have an integer number of wave periods $2 \pi/(k_T d)$  within the substrate of thickness $d$. Because the displacement component along $\hat z$ vanishes, so that these modes can normally not be detected using interferometry.

\subsubsection{ $\hat r\hat z$-polarized modes}
\label{r-z-polarized modes}

The solutions for  the modes polarized in the $\hat r-\hat z$ plane are a superpositions of longitudinal  and transverse modes polarized in this plane with amplitudes $B_e^\pm$ and $D_e^\pm$, respectively, and wave vector components $\pm k_T$ and $\pm k_L$ along z given by

\begin{eqnarray}
    k_L^2 &=&  \left(\frac{\omega}{v_L}\right)^2 - k_r^2, \quad  v_L=\sqrt{\frac{c_{11}}{\rho}}\\
    k_T^2 &=&  \left(\frac{\omega}{v_T}\right)^2 - k_r^2,  \quad v_T=\sqrt{\frac{c_{44}}{\rho}}.\nonumber
  \label{eqkLkT}
  \end{eqnarray}


Since the disk  has mirror-symmetry with respect to its  central  plane $z$=0, the solutions for the displacement field  can be classified according to whether they are mirror-symmetric [even, i.e., ${\bf u}(r, \theta, z)=-{\bf u}(r, \theta, z) $] or asymmetric [uneven, i.e., ${\bf u}(r, \theta, z)=+{\bf u}(r, \theta, z)$] with respect to reflection on this plane. In both cases, they must satisfy the following determinantal equation:

\begin{equation}
  Det = 
  \left[ 
    \frac
      {\tan \left(\frac{d k_T}{2}\right) }
      {\tan \left(\frac{d k_L}{2}\right) }
  \right]^{i_s} 
   +
      \frac{4 c_{44} k_L k_T k_r^2}
      {\left(k_T^2 - k_r^2\right) \left(c_{11} k_L^2+\left(c_{11}-2 c_{44}\right) k_r^2\right)} = 0.  
  \label{appendix:detSym}
\end{equation}
  
 \noindent Here, the exponent $i_s=1$ applies to symmetric modes (i.e., those for which $B_e^-=B_e^+=B_e$ and $D_e^-=-D_e^+=-D_e$) and $i_s=-1$ to the anti-symmetric modes ( $B_e^-=-B_e^+=-B_e$ and $D_e^-=+D_e^+=D_e$). The corresponding equation for one-directional Lamb wave propagation can be found in Ref.~\onlinecite{Auld90a} (Eqs. 10.18 and 10.19). The dispersion of the cylindrical Lamb modes  is shown in  Fig.~\ref{dispSymPlot} of the main text.

 \begin{figure}
  \centering
  \includegraphics[width=0.60\textwidth]{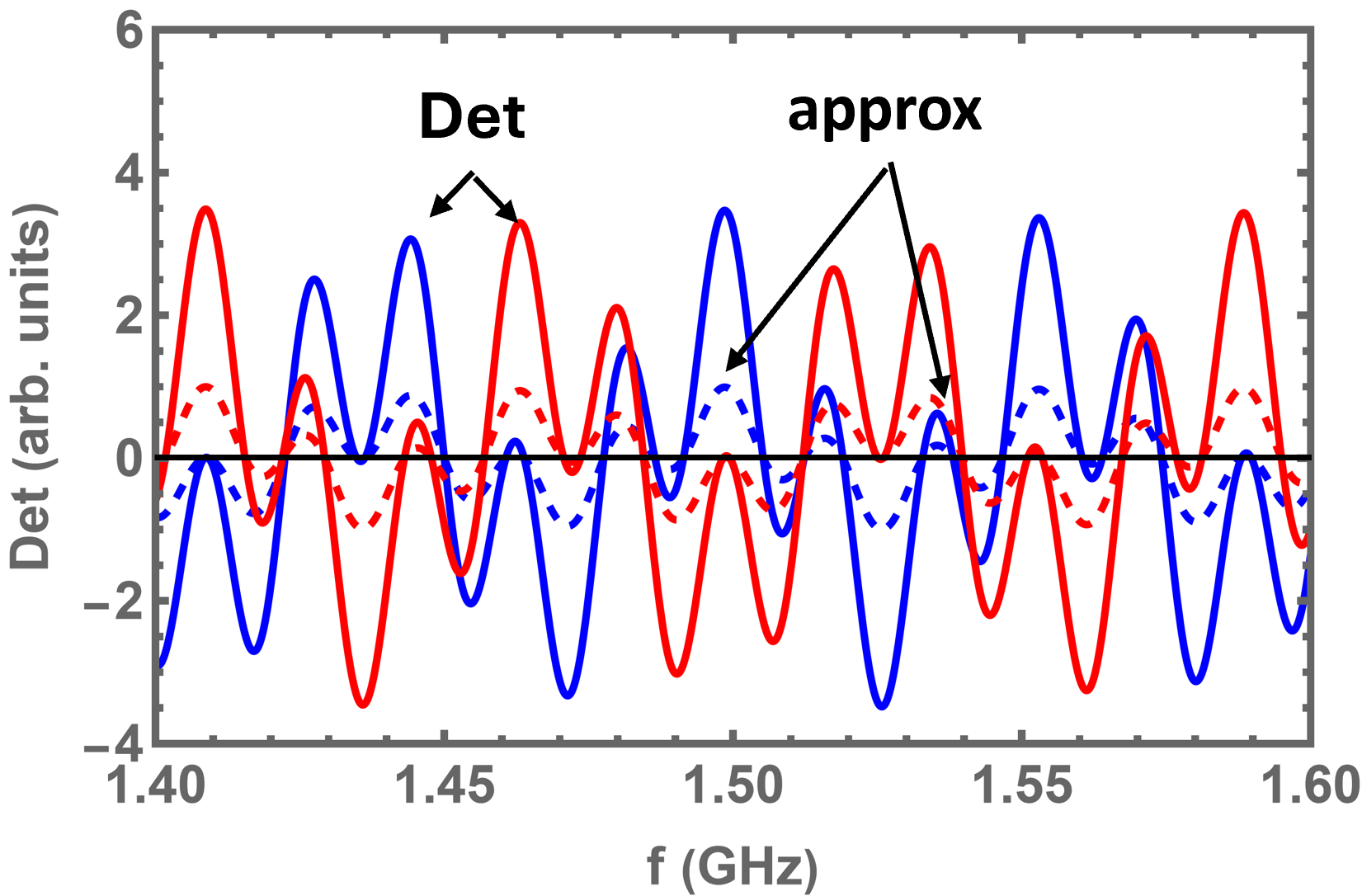}
  \caption{(solid line) Frequency dependence of the determinantal equation Eq.~\ref{appendix:detSym} of symmetric modes (blue, $is=1$) and anti-symmetric (red, $is=-1$). The Lamb wave solutions are those modes for within $Det = 0$.  The dashed line yields the approximation given by Eq.~\ref{detSymApprox}.} 
   \label{DetPlot}
\end{figure}

According to Eqs.~\ref{eqkLkT}, $k_L$ and $k_T$ are both real for small radial wave vectors in the range  $k_r < \omega/v_L $. For $\omega/v_L < k_r<\omega<v_T $, $k_L$ becomes purely imaginary: for a thick substrate, $\tan{k_Ld/2}\approx \rightarrow -\imath$. The longitudinal components of the displacement associated with the compressional wave thus become localized near the boundaries. 
Finally, for $k_r > \omega/v_T$  $k_T$ also becomes  purely imaginary and $\tan{k_T d/2}\approx \rightarrow -\imath$: the solutions in this case are surface acoustic waves (SAWs) bound to the surfaces. The extended (along $z$) solutions for $k_r<\omega/v_T$  consist of  several dispersion branches: for $k_r>\omega/v_T$, these solutions converge to  a single SAW mode per surface.

 \paragraph {Cylindrical SAW modes:}
  
 \cMM{DONE-MM: We seem to be starting a new list here.  Should there be a transition paragraph or a new section explaining the distinction between this list of solutions and the two types (a and b) ennumerated previously}Equation~\ref{detSym} in the main text and the expression derived below also apply for SAWs by making the $\tan$ ratio on the left side equal to one and the replacements $k_{T,L}\rightarrow \imath \alpha_{T,L}$, where $\alpha_{T,L}$ are real numbers. The dispersion of cylindric SAWs then becomes determined by the solutions of:

\begin{equation}
%
-4 v_T^4 \alpha_L \alpha_T k_{SAW}^2 + \left(\omega ^2-2 v_T^2 k_{SAW}^2\right)^2=0
\label{EqSAW}
\end{equation}

\noindent This dispersion differs from the one for SAWs propagating along a single direction  by the factor of 2 within the parenthesis (the dispersion for one-dimensional SAWs is given, e.g., by  Eq.~10.33 of \onlinecite{Auld90a}). In particular, the velocity of cylindrical SAWs (listed as 5761.66~m/s in Table~\ref{appendix:vTable} as calculated for the isotropic approximation of c-plane sapphire, cf. Table~\ref{appendix:pTable}) becomes \cPVS{Check with gsawnics!}{significantly lower} than for 1D-SAWs (6212.62~m/s).

\paragraph{Extended Lamb modes:}

The solid line in Fig.~\ref{DetPlot} displays the frequency dependence of the  determinantal equation Eq.~\ref{detSym} for symmetric modes (i.e., $is=1$) calculated for a sapphire substrate with a thickness $d=425~\mu$m. The Lamb wave solutions are those modes for which $Det =0$. These modes are displaced by frequencies $\Delta f_L =2d/v_L=13.6$~MHz corresponding to the inverse round trip of LA waves in the substrate. The solutions correspond very closely to the frequencies where the $\tan^{is}$ terms  in Eq.~\ref{detSym} vanishes, i.e., when

\begin{eqnarray}
 \cos{\left(\frac{k_L d}{2}\right)} \sin{\left(\frac{k_T d}{2}\right)}  &=& 0\\
 \cos{\left(\frac{k_R d}{2}\right)} \sin{\left(\frac{k_L d}{2}\right)}  &=& 0 \nonumber\\
\label{detSymApprox}
\end{eqnarray}

\noindent which is reproduced by the dashed line in the plot.
To a very good approximation, the solutions then consist of modes satisfying simultaneously $k_i d/\pi= m_i$ for  ($i=L,T$) with integers $m_i$, i.e., modes for which  an integer number of half cycles with length  $\pi/k_i$ fit within the substrate thickness $d$. This approximation may, however, fail at points of the dispersion where  dispersion modes anti-cross (see, Fig.~\ref{dispSymPlot} and the accompanying discussion in the main text). Under this approximation,
these solutions fall into two classes, depending on whether the disk surfaces correspond to a node or anti-node of waves with $r$ and $z$ polarization:


\begin{itemize}
\item {Modes with purely shear surface displacements:}
  
Symmetric modes with this character appear when $m_L$ and $m_T$ are both even, i.e.,  there is an even number of longitudinal and transverse half-cycle within the substrate. Equation~\ref{detSym} with $i_s=1$ is then satisfied with $\sin{(k_L d/2)}=\sin{(k_T d/2)}=0$. For the anti-symmetric modes, both $m_L$ and $m_T$ are non-even leading to $\cos{(k_L d/2)}=\cos{(k_T d/2)}=0$ in Eq.~\ref{detSym}. For both situations, the relation between the coefficients $B_e$ and $D_e$ is given by: 

  \begin{eqnarray}
    D_e&=&-\imath \frac{  c_{11} k_L^2 +  k_r^2 (c_{11} -2 c_{44})}{2 a c_{44} k_T k_r^2} B_e
  \end{eqnarray}
  

 \noindent The displacement amplitude reads:

 \begin{eqnarray}
 {\bf u}(r, \theta, z) &=&e^{i n \theta } B_e
 \left(
 \begin{array}{c}
  \frac{ (k_r r J_{n-1}(k_r r)-n J_n(k_r r))  \left(2 \cos \left(k_L z\right) c_{44} k_r^2+\cos \left(k_T z\right) \left(\left(c_{11}-2 c_{44}\right) k_r^2+c_{11} k_L^2\right)\right)}{k_r r^2 c_{44}} \\
  -\frac{ J_n(k_r r)  \left(2 \sin \left(k_L z\right) c_{44} k_L k_T-\sin \left(k_T z\right) \left(\left(c_{11}-2 c_{44}\right) k_r^2+c_{11} k_L^2\right)\right)}{c_{44} k_T} \\
 \end{array}
 \right) 
 \left( \begin{array}{c} {\hat r}\\ {\hat z}\\ \end{array} \right) 
 \label{uASym}
 \end{eqnarray}

   \noindent for $i_s=1$. 
The same expression, but exchanging $\sin\rightarrow \cos$  and $\cos\rightarrow \sin$ applies for the modes for $i_s=-1$. The surface displacement is restricted to the radial direction and  given by:

\begin{eqnarray}
{\bf u}\left(r, \theta, \frac{d}{2}\right) &=& B_e \frac{\omega ^2  (k_r r J_{n-1}(k_r r)-n J_n(k_r r))}{k_r r^2 {v_T}^2}  e^{i \theta  n} {\hat r}.
\label{uASymSurf}
\end{eqnarray}

\noindent Due to the absence of vertical (i.e., along $z$) surface displacements, these modes are not detectable by interferometry.

\begin{figure}[tbhp]
	\centering\includegraphics[width=0.75\textwidth, keepaspectratio=true]{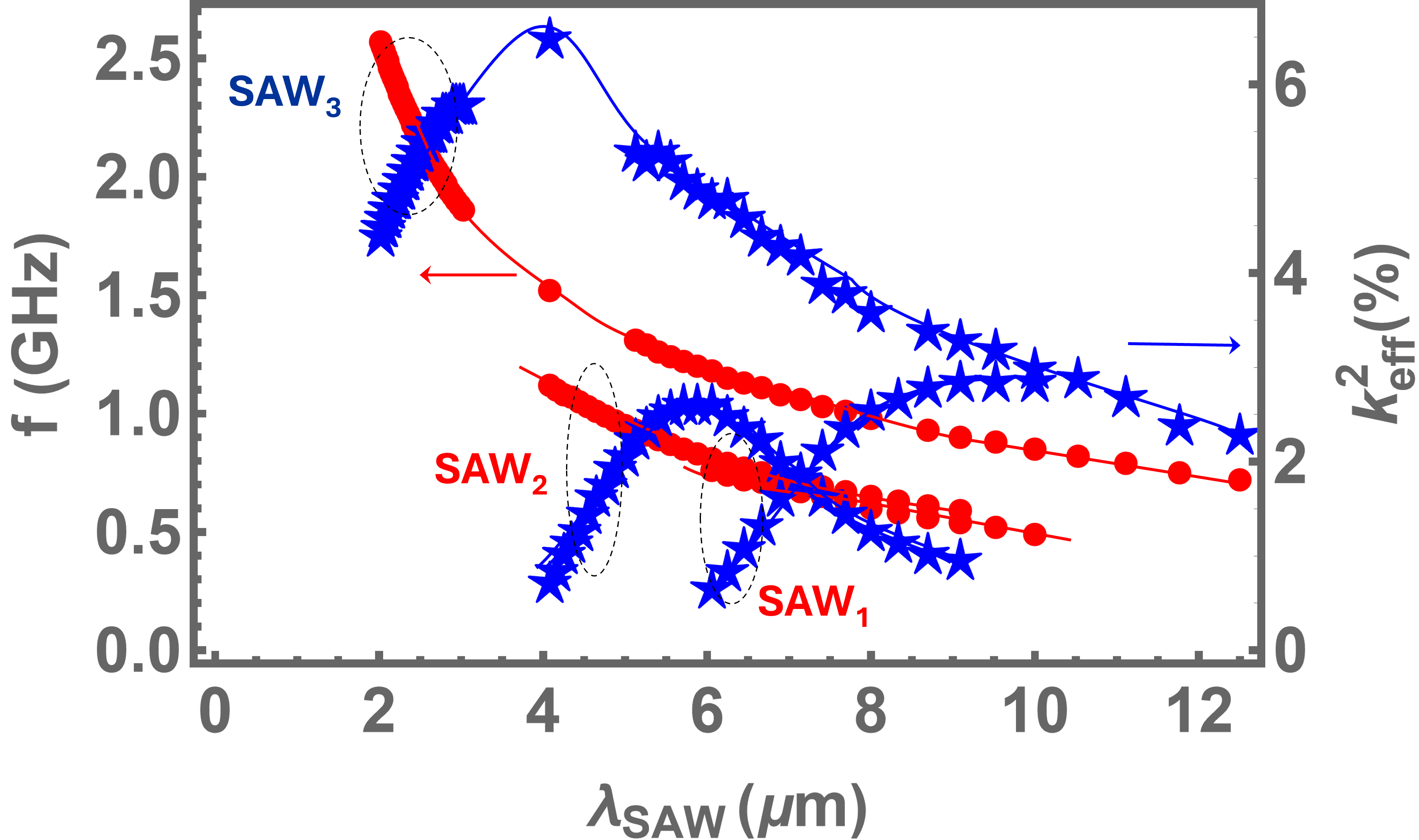}
 \caption{
 Calculated frequency (left vertical scale) and electromechanical coupling ($k^2_{eff}$ 
  as a function of the acoustic wavelength $\lSAW$ for piezoelectrically active SAW modes on c-plane sapphire coated with a 700~nm~thick ZnO layer.  The  calculations were carried out using the isotropic approximation for sapphire, cf. Table~\ref{appendix:pTable}. 
}
\label{appendix:SAWlambda}
\end{figure}

\item  {Modes with non-vanishing vertical surface displacements}
\label{Modes with non-vanishing vertical surface displacements}

For these modes, either $i_s=1$ and $m_L$ and $m_T$ uneven (leading to $\cos{(k_L d/2)}=\cos{(k_T d/2)}=0$ in Eq.~\ref{detSym}) or  $i_s=-1$ and both $m_L$ and $m_T$ even ($\sin{(k_L d/2)}=\sin{(k_T d/2)}=0$). The relation between the coefficients is given by:
 
  \begin{eqnarray}
    D_e&=&\frac{2 i  k_L}{a \left(k_T^2-k_r^2\right)} B_e
    \label{appendix:DeSym}
  \end{eqnarray}

The displacement field depends on the symmetry of the envelope functions according to Eq.~\ref{uSym} of the main text, which applies forfor $i_s=1$.

As in the previous case, the same expression, but exchanging $\sin\rightarrow \cos$  and $\cos\rightarrow \sin$ applies for the modes for $i_s=-1$. The radial displacement vanishes  at the surface while the vertical (i.e., along $z$)  displacement becomes:

\begin{eqnarray}
{\bf u}\left(r, \theta, \frac{d}{2}\right) &=& 
 B_e  {k_L}   
\frac{ \omega ^2   }{\omega ^2-2 k_r^2 v_T^2}   J_n(k_r r)   e^{i n \theta } {\hat z}.
\label{uSymSurf}
\end{eqnarray}

These modes are of particular interest due to their $z$-polarization near the surface, which enables detection by interferometry as well as efficient piezoelectric excitation by BAWRs. Here, one needs to consider that the strongest piezoelectric stress $\sigma_{zz}$ exert by the ZnO film is mediated by the piezoelectric strain coefficient $e_{33}$, which couples the vertical rf-electric field in-between the BAWR electrodes to the strain along z according to $\sigma_{zz} = e_{33} \partial u_z/\partial z$.

The surface displacement amplitude for these modes given by Eq.~\ref{uSymSurf} resonates at 

\begin{equation}
\omega = \sqrt{2} v_T k_r, 
\label{EqvTsqrt}
\end{equation}

\noindent the so-called Lam{\'e} wave solution~\cite{Auld90a}. These modes are fully z-polarized at the surface and have  a linear dispersion with a propagation velocity  along the $r$-direction   $\sqrt{2}$ higher than the transverse  velocity. 

\end{itemize}

\begin{figure}[tbhp]
	\centering\includegraphics[width=0.45\textwidth, keepaspectratio=true]{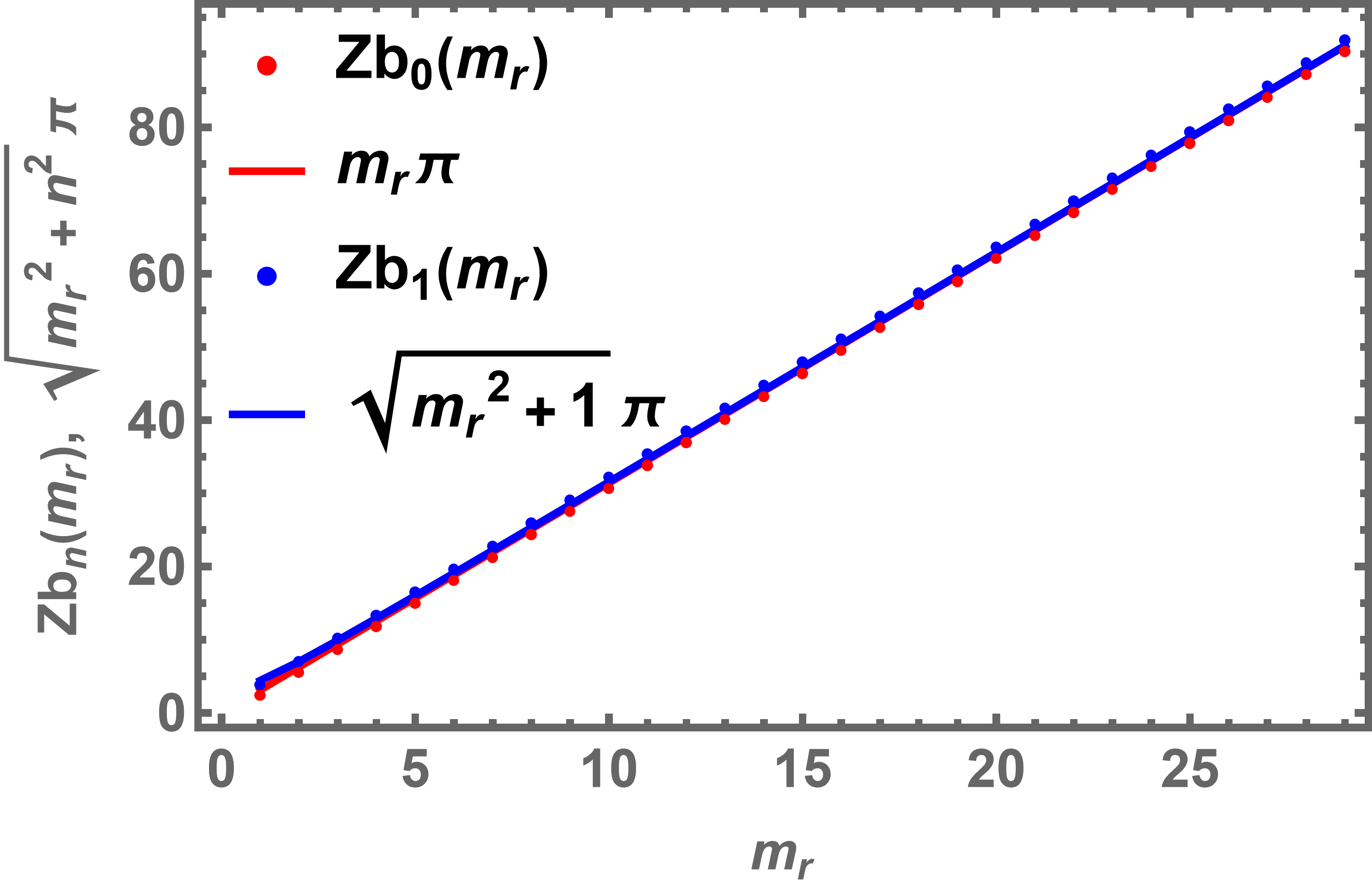}
 \caption{
  Comparison betwen $\sqrt{m^2_r+n^2} \pi$ and $Z_{b_n}(m_r)$ for different winding numbers $n$ and radial confinement indices $m_r$.   
}
\label{appendix:lambApprox}
\end{figure}

\subsection{SAW modes in ZnO-coated c-plane sapphire}
\label{SAW modes in ZnO-coated c-plane sapphire}

The presence of the textured ZnO layer, which has an acoustic velocity substantially lower than the substrate, can induce different SAW modes with phase velocities $\vSAW$ and piezoelectric generation efficiency depending on the ratio between the SAW wavelength ($\lSAW$) and the ZnO thickness.

Figure~\ref{appendix:SAWlambda} displays dependence of the SAW frequency $\fSAW=\vSAW/\lSAW$ and electromechanical generation efficiency $\keff$ on $\lSAW$ as calculated for a c-plane sapphire substrate coated with $\dfilm= 700$~nm-thick ZnO film. The calculations were carried out for a 1D model for SAW propagation using the elastic properties listed in Table~\ref{appendix:pTable}. Due to the much smaller acoustic velocities  of ZnO as compared to sapphire, the ZnO film acts as an acoustic waveguide supporting different transverse modes denoted in the figure as SAW$_i$, $i=1,2,3$. For simplicity, we include in the figure only modes with $\keff>0.25$, which can be efficiently excited by the IDT. Note that $SAW_3$ has the highest $\keff$ as well as the highest acoustic velocity.


\subsection{Solution for a disk of radius {$\rfilm$}}
\label{Solution for a disk of radius a}

As stated in the main text, approximate solutions for Lamb modes in a disk of finite radius $\rfilm$ are obtained by imposing the constraint on the radial wave vector given by Eq.~\ref{EqZB}. To a very good approximation,  $Z_{b_n}(m_r) \approx m_r \pi$. This assumption is confirmed the comparison of these two quantities as a function of $m_r$ displayed in  Fig.~\ref{appendix:lambApprox}.

\section{Radio-Frequency phase shifter}
\label{Radio-Frequency phase shifter}

\begin{figure}[tbhp]
	\centering \includegraphics[width=0.7\textwidth]{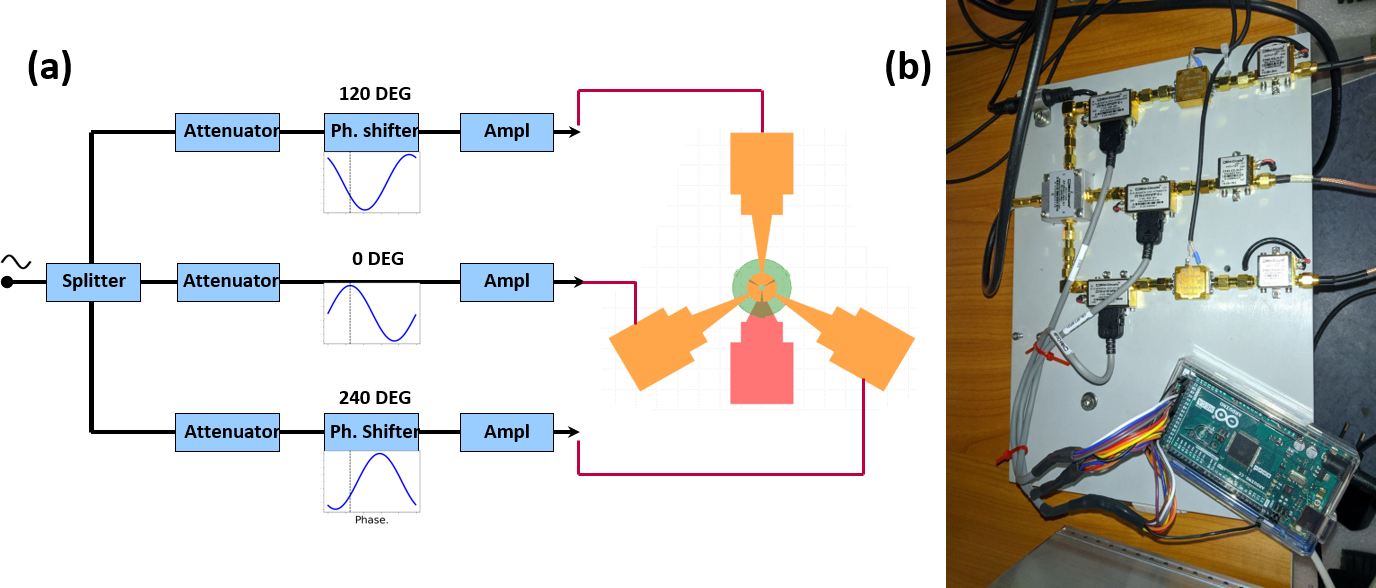}
\caption{Tunable phase shifter for the excitation of helical acoustic waves. (a) The input rf-signal is split by a passive 3-fold splitter into three components, which are shifted in phase by electrically controlled phase shifters. The attenuator and amplifier allow the control of the amplitude of the rf-drives to be applied to the sector BAWRs. (b) Photograph of the setup.}
  \label{PhaseShifter}
\end{figure}

Figure~\ref{PhaseShifter}(a) displays the schematic diagram of a programmable rf phase shifter used for the excitation of helical waves. In the setup, the input rf drive is first split by a passive 3-fold splitter into three components. These are then shifted in phase relative to each other by two electrically controlled phase shifters. The programable attenuators and amplifiers  in each of the three arms enable allow the control of the amplitude of the rf-drives applied to each  sector BAWR. Panel (b) of Fig.~\ref{PhaseShifter}(b) shows a photograph of the setup.

\section{Videos of helical acoustic modes}
\label{Displacement_Videos}

The following mp4 video files display animations of helical acoustic fields: 
\begin{itemize}
\item Clockwise-rotating helical acoustic field (video SM-Video-CW-240208-001\_F2.mp4): time evolution of a clockwise rotating field at a frequency of 1.5033 GHz (cf. Figs.~\ref{Fig4}(b)-(d) of the main text).
\item Counterclockwise-rotating helical acoustic field (video SM-Video-CCW-240206-013\_F2.mp4): time evolution of an counterclockwise rotating field at a frequency of 1.5033 GHz (cf. Figs.~\ref{Fig4}(f)-(h) of the main text).
\end{itemize}

The videos were elaborated from the measured phase-resolved interferometric profiles by first subtracting the unstructure background signal and then adding a fixed phase shift per frame (40 frames for a phase span of 360$^\circ$).  

\end{document}